\documentclass[singlecolumn, showpacs, secnumarabic,amssymb,nobibnotes,aps,prl]{revtex4-2}
\usepackage{graphicx, graphics, amsmath, amssymb, rotate, textcomp, gensymb}
\usepackage{mathrsfs, float}
\usepackage[T1]{fontenc}
\usepackage{dcolumn}
\usepackage{physics}
\usepackage{wrapfig}
\usepackage{xcolor}
\usepackage[none]{hyphenat}
\usepackage{url,hyperref}
\usepackage{times}
\usepackage{multirow}
\begin{document}

\title{Quicker flocking in aligning active matters for noisier beginning}

\author{Sohini Chatterjee}\thanks{Equal contributions}
\author{Sohom Das}\thanks{Equal contributions}
\author{Purnendu Pathak}\thanks{Equal contributions}
\author{Tanay Paul}\thanks{Equal contributions}
\author{Subir K. Das}
\email{Email of corresponding author: das@jncasr.ac.in}
\affiliation{Theoretical Sciences Unit and School of Advanced Materials, Jawaharlal Nehru Centre for Advanced Scientific Research, Jakkur, Bangalore 560064, India}
\date{\today}

\begin{abstract}
The constituents in a class of active matter systems change their directions of motion by being influenced by the velocities of the neighbors. 
Such systems may undergo phase transitions, with respect to ordering in the velocity field, as well as clustering in the density field, when the strength of an externally imposed noise is varied. 
Via computer simulations, with a well-known model, that faithfully represents these systems, we show that evolutions in both clustering and ordering exhibit certain interesting features that were hitherto unrealized. 
The transformations occur quicker, following quenches to a fixed final state, below the transition point, for disordered starting states that are farther away from the 
``critical" noise strength.  
This implies earliest arrival of the farthest, at a given destination. 
Detailed analysis of the results, combined with the outcomes from a similar study of para- to ferromagnetic transitions, show that the variation in critical fluctuations in the initial configurations can lead to such interesting effect. 
We quantify this via the Ornstein-Zernike theory. 
\end{abstract}

\maketitle

\section{Introduction}
\sloppy

Evolution towards a new state, following the quench of a homogeneous system inside the ordered or coexistence region of a phase diagram, is a complex dynamical process \cite{bray_article, onukibook}.
Understanding of such kinetics of phase transitions is of much interest for passive as well as active matters \cite{bray_article, puribook, Cates2013, Bhaskaran2013, Rajesh2012, das14, das2017, paul_bera, puricoarsening, TP}.
For the quenching experiments, traditionally one considers initial configurations that possess constituents having random directionalities or positions \cite{das2017,puricoarsening,lowen_pre,activekinetic,experiment0,experiment1,experiment2,experiment3,TP,bera,mishra,bhaskar,chate,paul_bera,paul2022,miclusters}.
These correspond to states with insignificant spatial correlations \cite{fisher_theory}. 
However, it is important to carry out similar exercises with initial states of different varieties.
Such practical protocol may lead to interesting outcomes when there exists interplay \cite{Bray_Tc,Cugliandolo_Tc,saikat_EPJB,Das_JPCM,koyelina_PRE} between critical \cite{onukibook,fisher_theory} and coarsening phenomena \cite{bray_article,onukibook,puribook}.
In the long time ($t$) limit, it is of interest to see \cite{Bray_Tc,Cugliandolo_Tc,saikat_EPJB,Das_JPCM,koyelina_PRE} if universal pictures emerge -- whether the results remain same irrespective of the degree of spatial correlations at the starting configurations. 
Furthermore, it is important to investigate how the change in (steady-state) correlation, with the shift of the initial state towards a ``critical" point, may affect the nature and alter the longevity of the pre-asymptotic coarsening process \cite{skdLang}.
Even if the asymptotic rate of growth remains unchanged, despite the variation in the starting states, there will be differences in overall evolution.
Quantification and understanding of this bear much practical and fundamental relevance in vast varieties of physical, chemical as well as biological systems. 

Here, for a type of active matter \cite{vicsek95, vicsek1997, Vicsek_rep2012, das14, das2017}, we study kinetics of transitions in two different fields, viz., ordering in the velocity field and clustering in the density field.
The aim is to investigate the change in the rates of evolutions with the variation of initial state. 
It appears that for both the fields the evolutions get slower as the initial state is chosen closer to the critical point of transition \cite{fisher_theory, stanleybook}. 
In thermodynamic sense, this implies that the farthest arrives at a destination earliest, which is counter-intuitive \cite{mpemba,skdLang}. 
Similar effect, in the context of passive matter has become a topic of considerable interest \cite{aristotle,skdLang, water_arxiv, bechhoffer, jeng, baity, nv, schat_potts, gal_raz, chaddah, avinash, chetrite, lasanta, torrente, mompo, rajesh, biswas, jin, tao, burridge, lu_raz, klich, auerbach, vynnycky, xi, tang, lowen22, ahn, greaney, pemartin21,mpemba, Hayakawa2023}.
E.g., faster freezing of a hotter sample of liquid water \cite{mpemba}, than a colder one, when quenched to a fixed subzero temperature, has drawn significant attention. 
In the latter case, recent studies suggest that metastability could be the reason behind the surprising outcome \cite{water_arxiv,rajeshmpemba,bechhoffer,jin}.
Our study, in addition to identifying a counterpart in active matter phase transitions, provides further new insight that metastability is not a necessity for such observation \cite{water_arxiv,rajeshmpemba}. 
This conclusion is further strengthened by presenting similar results for para- to ferromagnetic (PF) transitions \cite{nv,schat_potts}.
The results from multiple types of transformations point towards the existence of universality that we quantify via the Ornstein-Zernike theory \cite{stanleybook}.

\section{Model and Methods}

We consider a model \cite{vicsek95, vicsek1997, gregoire04, albano08, albano2009} which incorporates a dynamical alignment interaction amongst the constituent point-like active particles. 
These particles self-propel on off-lattice planes with constant speed $v_0$. 
The model exhibits rich structural and dynamical features that arise from the competition between this alignment interaction and a random noise $\zeta$, that plays role analogous to temperature ($T$) in the PF transition.
While moving with a velocity $\vec{v}_j ~(= v_0 e^{i \theta_j})$, at time $t$, after a step of size $\Delta t$, a particle $j$ can change its direction $\theta_j$, being influenced by the average direction of motion of its neighboring particles, with an uncertainty due to $\zeta$. 
Thus \cite{vicsek95},
    $\theta_j (t+\Delta t) = \langle \theta (t) \rangle _{ \mathscr{R}_j } + \zeta$, 
$\mathscr{R}_j$ defining the neighborhood around the $j$-th particle, $\langle \theta (t) \rangle _{ \mathscr{R}_j }$ being calculated as \cite{vicsek1997} 
    $\langle \theta (t) \rangle _{ \mathscr{R}_j } = {\rm arg} {[ \sum_{k \in \mathscr{R}_j} e^{i \theta_k (t)}]}$.
We treat $\zeta$ as a uniform random noise within $[-\eta/2, \eta/2]$, $\eta \in [0,2\pi]$. 
The update of position $\vec{r}_j$, for particle $j$, is given by \cite{vicsek95, albano2009}
    $\vec{r}_j (t+\Delta t) = \vec{r}_j (t) + \vec{v}_j (t+\Delta t) \Delta t$.
We set $\Delta{t}=1$.
The transition from a disordered to a flocking state, in this model, to be referred to as the Vicsek Model (VM), below $\eta_c$, the critical noise strength, is  characterized via the calculation of the order parameter \cite{vicsek95}
    $v_a = N^{-1}| \sum_{j=1}^{N} e^{i \theta_j}|$, 
$N$ being the total number of particles in a periodic box of size $L \times L$. 
The average value of $v_{a}$ is expected to be finite in a flocking state, and vanishingly small in a disordered one.

For the magnetic transition we have performed Monte Carlo (MC) simulations with the Ising model (IM), having the Hamiltonian \cite{onukibook, puribook, landaubinderbook}
    $H=-J\sum_{<i,j>}S_iS_j$,
$S_{i(j)}(=\pm1)$ being a spin or an atomic magnet sitting at a lattice site $i (j)$, with $J$ being the strength of the interaction between two such nearest neighbors. 
For a thermodynamically large lattice size, i.e., for $L=\infty$, in space dimension $d=2$, this model exhibits a phase transition \cite{onukibook, puribook, landaubinderbook} at the critical temperature $T_c\simeq 2.269 J/k_B$, $k_B$ being the Boltzmann constant.
In our MC simulations, a trial move is made by randomly choosing a spin and altering its sign \cite{landaubinderbook}. 
The move is accepted via the Metropolis algorithm \cite{landaubinderbook}. 
$L^2$ such trials make a MC step (MCS), our time unit.
For this model the equilibrium order parameter $m$ $(= \sum_i{S_i}/L^2)$ is the magnetization per spin.

For the VM, alongside a transition in the velocity field, a flavor of vapor-liquid transition is also present.
Such phase separation can be probed via the two-point equal-time correlation function \cite{bray_article}
    $C(r,t)=\langle\psi(\vec{r},t)\psi(\vec{0},t)\rangle-\langle\psi(\vec{r},t)\rangle\langle\psi(\vec{0},t)\rangle; ~~~ 
    r=|\vec{r}|$.
Here $\psi(\vec{r},t)$ is the local density-field order-parameter at time $t$ at a space point $\vec{r}$ on a lattice to which our off-lattice systems are mapped on, for the sake of convenience \cite{roydas}. 
We have assigned $\psi=+1$ if the local density, computed within a circular region of unit radius around the point $\vec{r}$, is larger than the overall particle density, $\rho_{0}=N/L^{2}$, otherwise $-1$.  
Thus, for the analysis purpose, we have a picture analogous to the IM \cite{majumder11}.
The enhancement of density inhomogeneity in a system can be quantified by extracting 
a length scale, $\ell(t)$, from the decay of $C(r,t)$ to a reference value $C_{\rm {ref}}$, at different times. 
Analogously, depending upon the density, it may also be instructive to calculate average number of particles in (discrete) clusters after appropriate identification of the latter \cite{roydas,paul_bera} by setting an interparticle distance ($d_c)$ criterion for particles belonging to a common cluster. 
Here we will present results for the largest clusters, as a function of time, which also is an important quantity.   
In this study, we have considered $\rho_0=0.5$, $v_0=0.03$, $C_{\rm{ref}}=0.1$ and $d_c=1.5$. 
The choices for first two parameters provide \cite{vicsek1997} $\eta_c \simeq 1.5$, when $L=\infty$.
Initial or starting configurations for the VM, with almost vanishing $v_a$, have been prepared by carrying out simulations at different $\eta$ values, 
$\eta_s (>\eta_c)$, beginning with configurations having random assignments of both positions and velocities. 
The steady-state configurations, thus obtained, have been quenched to a fixed noise strength $\eta_{_f}=0.5$.

Each simulation box, for the VM, has been divided into square cells of unit side length. 
For a particle of interest, the alignment interaction range $\mathscr{R}$ is 
defined as the cell in which the particle is located, along with its eight neighboring cells. 
Results are presented for systems of size $L=256$, with $N=32768$. 
The system size for the IM remains the same, containing 65536 spins.
In this case, for the preparation of initial configurations we have chosen starting temperatures $T_s$ above $T_c$.
Configurations from there, with small $m$, are quenched to a final temperature $T_f=0.5T_c$. 
Unless otherwise mentioned, all discussions are for the VM. 
The quantitative results correspond to statistics over $20000$ initial configurations.

\section{Results}

\begin{figure}[]
    \centering
    \includegraphics[width=0.48\textwidth]{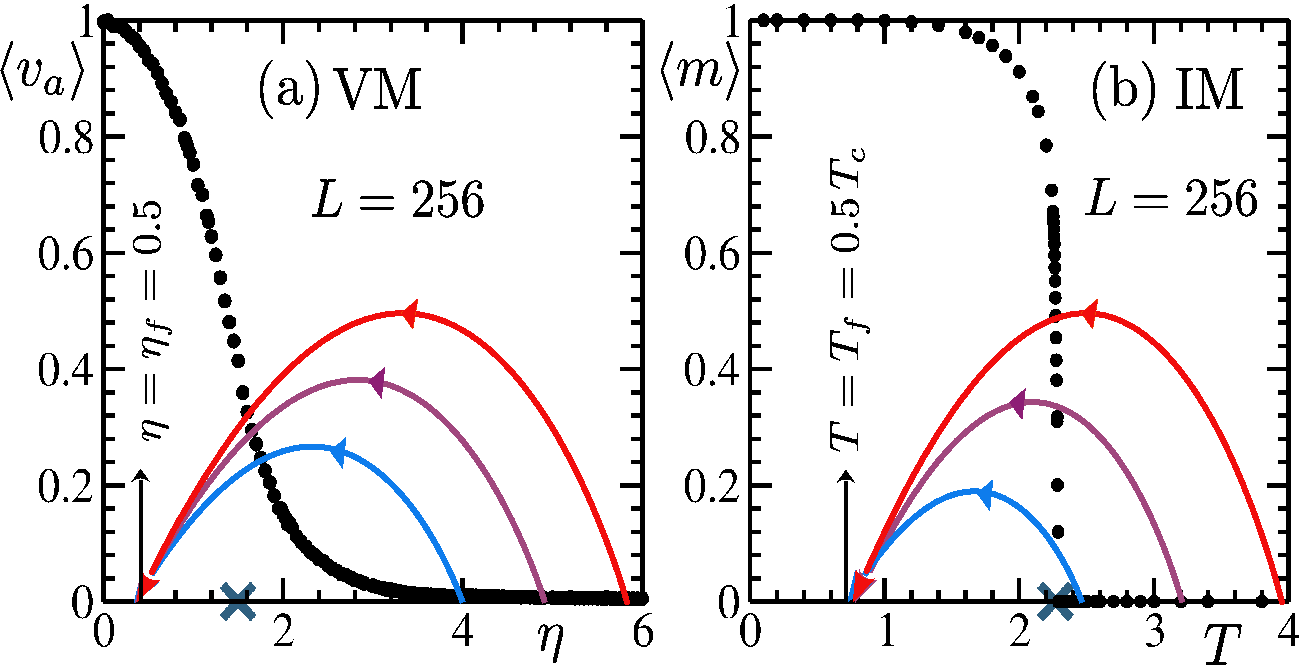}
    \caption{(a) Average order parameter for the active matter model is plotted versus $\eta$. The protocol for quenches from different starting noise strengths, $\eta_s$, to a final strength, $\eta_{_f} (= 0.5)$, has been sketched.
    (b) Same as (a) but here the phase diagram and the protocol (for quenches from different starting temperatures $T_s$ to the specified $T_f$) are for the magnetic model.
    The critical points in the thermodynamic limit are marked by the crosses \cite{vicsek1997,fisher_theory}.}
    \label{fig:PhaseDiagram}
\end{figure}

In Fig. \ref{fig:PhaseDiagram} (a) the symbols show a plot of the average order parameter, $\langle v_a \rangle$, versus $\eta$, for the VM.
For each $\eta$, $\langle v_a \rangle$ was obtained by constructing distributions, via many simulations, starting with random initial configurations, over long times, in the respective steady state. 
The curved lines, with arrows, depict the protocol we have used for the investigation of the initial state dependence of coarsening that may lead to the effect discussed above. 
Analogous phase diagram, in magnetization versus $T$ plane, along with the protocol, for the IM, is shown in Fig. \ref{fig:PhaseDiagram} (b).

\begin{figure}[]
    \centering
    \includegraphics[width=0.48\textwidth]{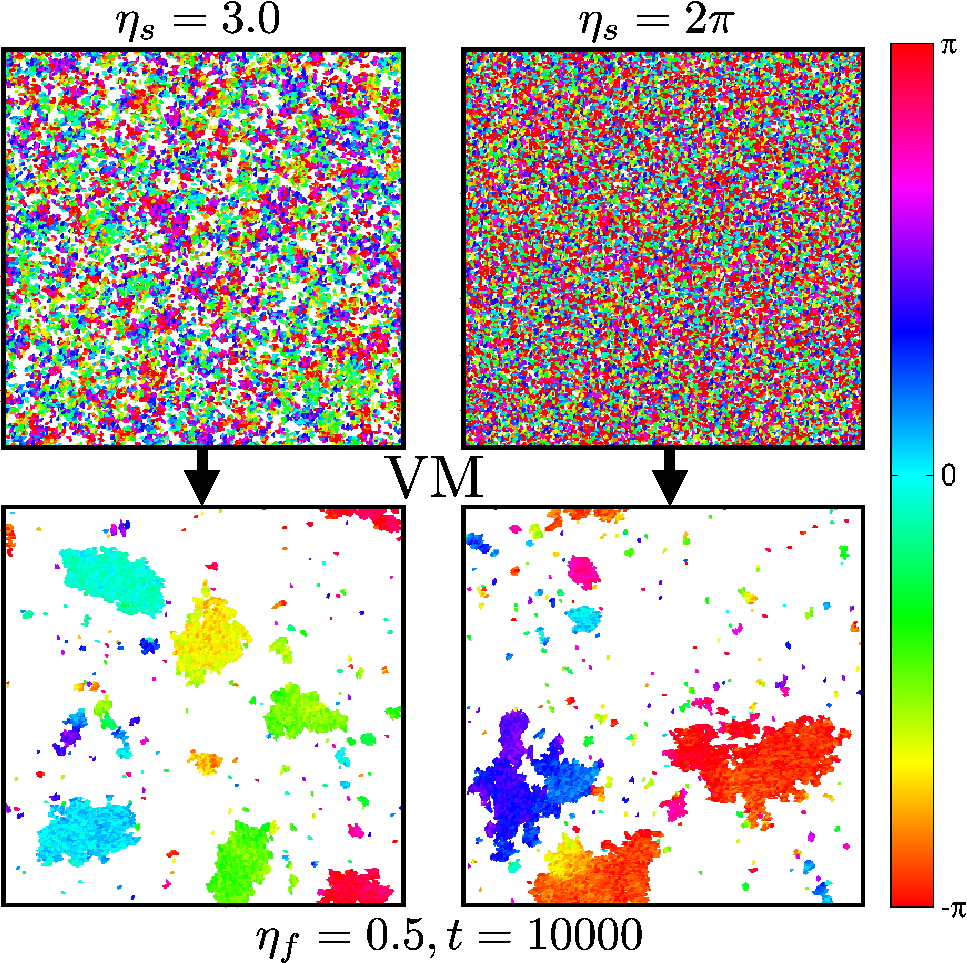}
    \caption{Upper frames: Representative steady-state snapshots for two different values of $\eta_s$. Lower frames: Snapshots from $t=10000$, taken during evolutions, following the quenches of the upper configurations to $\eta_{_f}=0.5$. The positions of the particles are marked. The color coding represents directionality of motion.}
    \label{fig:Snaps_both_eta}
\end{figure}

In Fig. \ref{fig:Snaps_both_eta} we show certain snapshots for the VM. 
The two upper frames contain representative steady-state pictures from two values of $\eta_s$. 
For the higher $\eta_s$, there exists more randomness in both positions and velocities. 
The lower frames display evolution snapshots that are obtained following quenches of the upper snapshots to $\eta_{_f} = 0.5$. 
Clearly, both velocity ordering and density field clustering are occurring quicker for $\eta_s = 2\pi$, i.e., for the starting configuration that is farther from $\eta_c$. 
Note that for both the $\eta_s$, the snapshots are recorded at the same instant following the quenches.
This difference in the pace of phase transitions is not only a fact corresponding to this particular combination of initial configurations, the trend is rather general.

\begin{figure}[]
    \centering
    \includegraphics[width=0.48\textwidth]{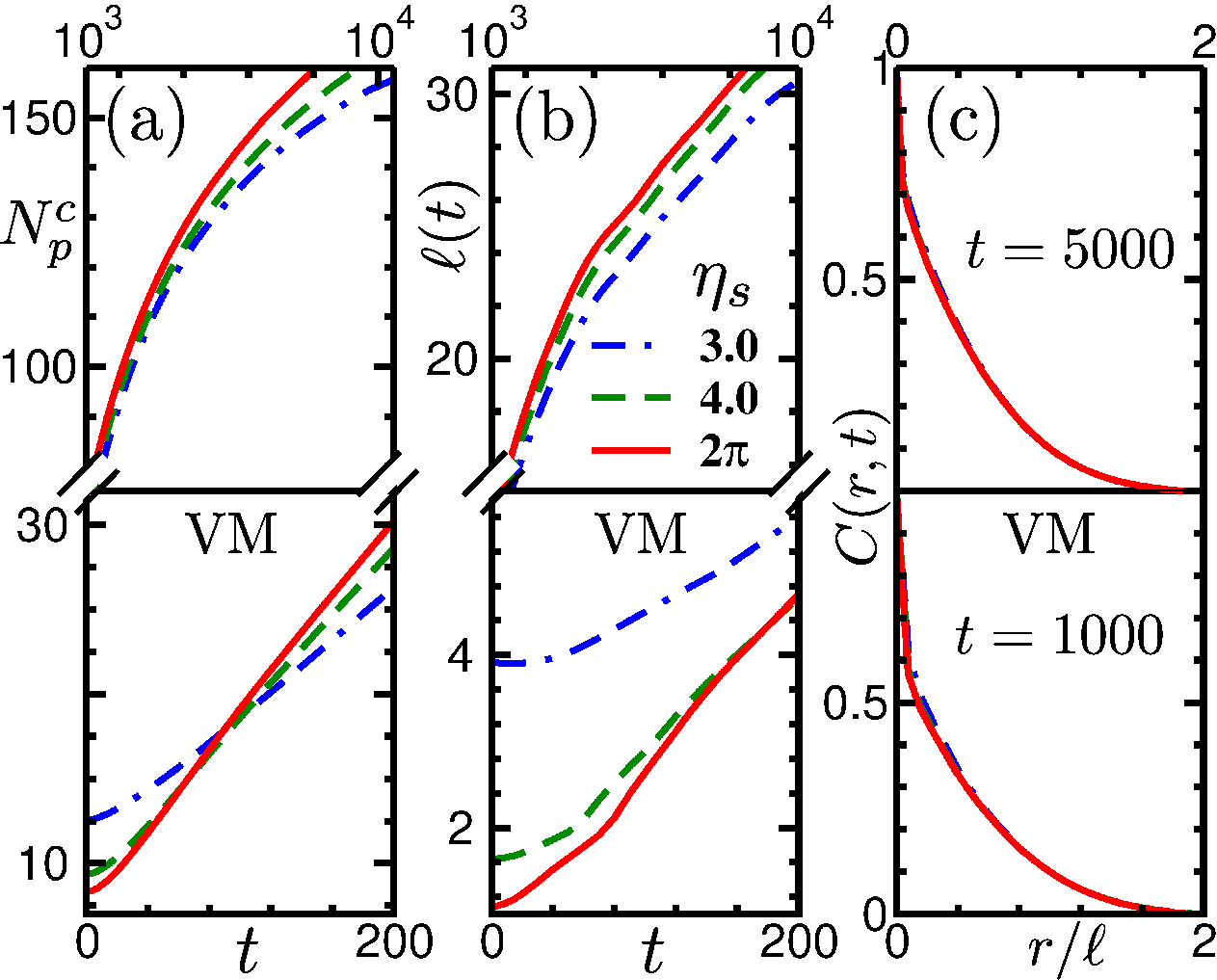}
    \caption{(a) Plots of (averages of the) largest cluster sizes, related to the density field phase separation, versus time, for $\eta_s =2\pi,4.0,$ and $3.0$. (b) Plots of the corresponding average domain lengths, versus time. (c) Scaling plots of two-point equal-time density correlation functions for the same set of $\eta_s$. The results correspond to $\eta_{_f} = 0.5$.}
    \label{fig:data}
\end{figure}

In Fig. \ref{fig:data} (a) we show the average sizes of largest clusters, versus time, for three $\eta_s$. We have divided the results into two time regimes.
The lower plots represent data from early times and the later time data are shown in the upper half of the broken frame. 
The orders of appearances of the data sets are different for very early and the later time regimes. 
Clearly, at late times the clusters are larger for higher starting noises. 
In the literature of phase-separation kinetics, it is, however, more customary to discuss the time dependence of $\ell$. 
In Fig. \ref{fig:data} (b) we show $\ell$ versus $t$ plots for the same set of $\eta_s$. 
This also conveys same surprizing message.
One may inquire whether there exists structural similarity during evolutions for all the considered values of $\eta_s$.
To verify this, in Fig. \ref{fig:data} (c) we show the scaled correlation functions \cite{bray_article}, the lower and upper sub-frames containing data from early and late times.
Nice collapse of data from each $\eta_s$, upon scaling of the distance by corresponding $\ell$,
confirms the existence of structural similarity, though at somewhat late times.

\begin{figure}[]
    \centering
    \includegraphics[width=0.48\textwidth]{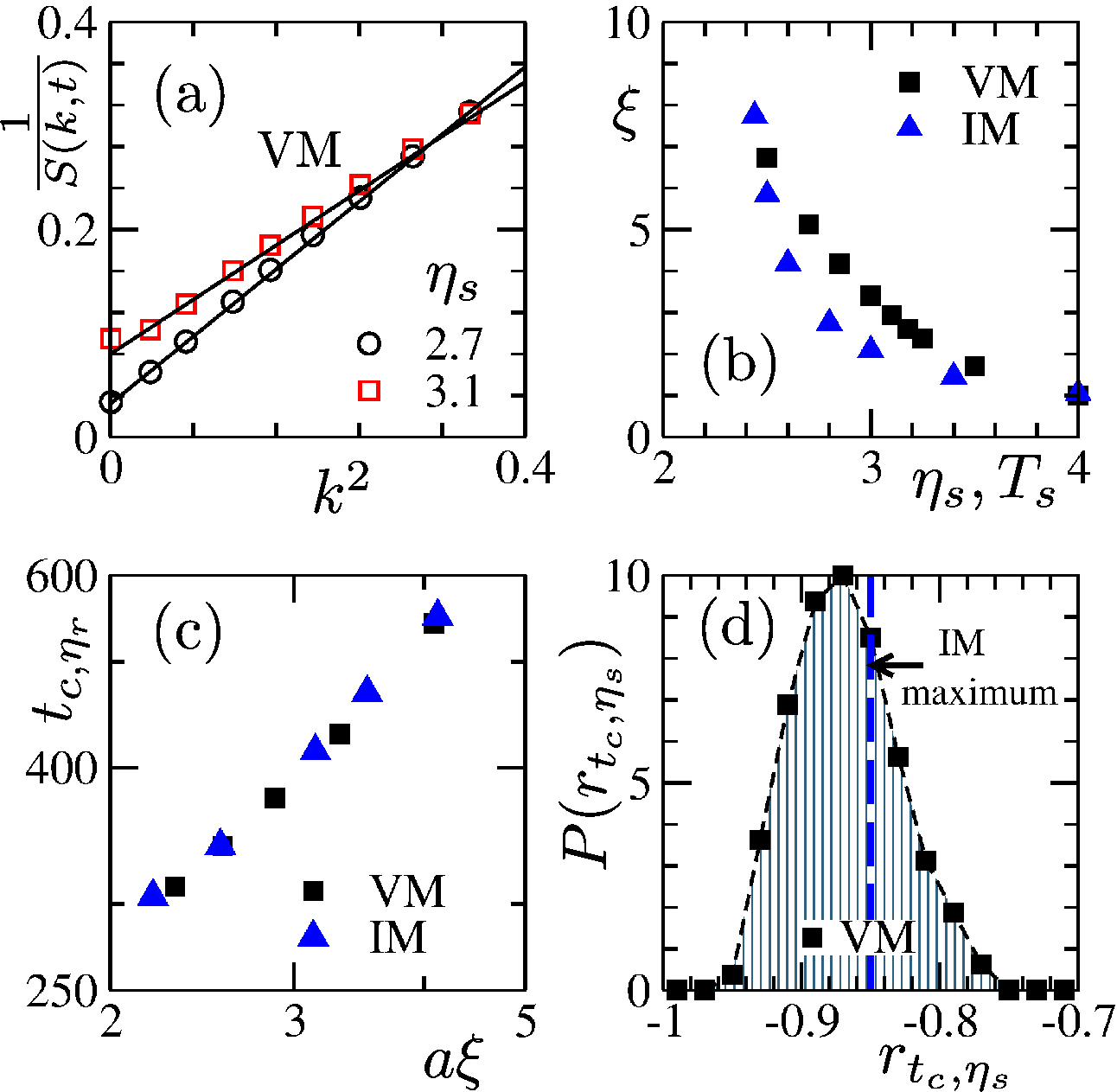}
    \caption{(a)  Plots of $1/S(k)$, versus $k^2$, for two values of $\eta_s$. The continuous lines correspond to Ornstein-Zernike (OZ) expectation. (b) Correlation lengths $\xi$, calculated using the OZ relation, are plotted against $\eta_s$ and $T_s$, respectively, for the VM and the IM. (c) Plots of crossing times $t_{c,\eta_r}$, versus $\xi$, for the VM and the IM. To compare the scaling for the two models, a prefactor $a$ has been introduced in the abscissa. (d) Distribution of the Pearson correlation coefficient, $r_{t_c,\eta_s}$, for the VM. The location of the corresponding maximum for the IM has been shown by the dashed vertical line.}
    \label{fig:xi_tc}
\end{figure}

For spin systems \cite{nv, schat_potts}, similar effect was argued to arise due to differences in spatial correlations in the initial states.
With the intention of examining whether such a scenario applies here as well, we calculate the structure factor, $S(k)= \langle\psi_k\,\psi_{-k}\rangle$, $\psi_k$ being the Fourier transform of $\psi(\vec{r})$, to estimate $\xi$, the correlation length, via the Ornstein-Zernike (OZ) relation $1/S(k) \propto (1 + k^2\xi^2)$.
In Fig. \ref{fig:xi_tc} (a) we show, for the VM, plots of $S(k)^{-1}$, versus $k^2$. 
The trend, with the variation of $\eta_s$, reflects the fact that fluctuation becomes critical with the decrease of $\eta_s$.
The linear appearances imply that the OZ behavior is obeyed. From the slopes we calculate $\xi$. 
In Fig. \ref{fig:xi_tc} (b) we show $\xi$, versus $\eta_s$.
There we have also plotted IM data, versus $T_s$. 
Critical enhancements are very clear. 

Next, we calculate the times, $t_{c,\eta_r}$, for crossings of the domain length plot for a reference value of $\eta_r=\eta_s=2\pi$, 
the highest admissible noise within the model, with those for others.
For the Ising case we represent the crossing times as $t_{c,T_r}$, with $T_r=\infty$, that corresponds to a perfectly random configuration.
In Fig. \ref{fig:xi_tc} (c) we plot the corresponding results for both the models, versus $\xi$ (at $\eta_s$ or $T_s$).
Nice (almost) linear scaling for both the models implies possible existence of a universal feature.
Note that these plots also imply that systems with high initial noise or temperature eventually overtake all other systems starting with corresponding lower numbers.

All our presented results are obtained by averaging over runs with large number of independent initial configurations, to draw the conclusions carefully.
Nevertheless, to further ascertain the correctness of these, given that the observation is counterintuitive, we have calculated the Pearson correlation coefficient, $r_{t_c,\eta_s}$, defined as \cite{pearson95}
\(r_{t_c,\eta_s} = \sum_{s=1}^{n} x_s y_s / [(\sum_{s=1}^{n} x_s^2) (\sum_{s=1}^{n} y_s^2)]^{1/2}\),
where $x_s= t_{c,\eta_r}(\eta_s) - \bar{t}_{c,\eta_r}$ and $y_s = \eta_s - \bar{\eta}_s$, with $\bar{t}_{c,\eta_r}$ and $\bar{\eta}_s$ being the average crossing time and the average noise strength, respectively, for a sample size $n=3$, corresponding to $\eta_s=2.5,3,4$.
For this purpose, we have divided the full sets of initial configurations into $400$ subsets. 
For each of these subsets, $r_{t_c,\eta_s}$ is calculated. 
From these we have derived $P(r_{t_c,\eta_s})$, the probability distribution for $r_{t_c,\eta_s}$. The distribution for VM is shown in Fig. \ref{fig:xi_tc} (d). 
Quite clearly the peak appears very close to $-1$, implying anticorrelations as clear likelihood.
This is because higher randomness, in each small subset of initial configurations, leads to quicker clustering, i.e., the farthest arrives the earliest.
The location of corresponding maximum for the IM is marked.

\begin{figure}[]
    \centering
    \includegraphics[width=0.65\textwidth,  height=0.3\textwidth]{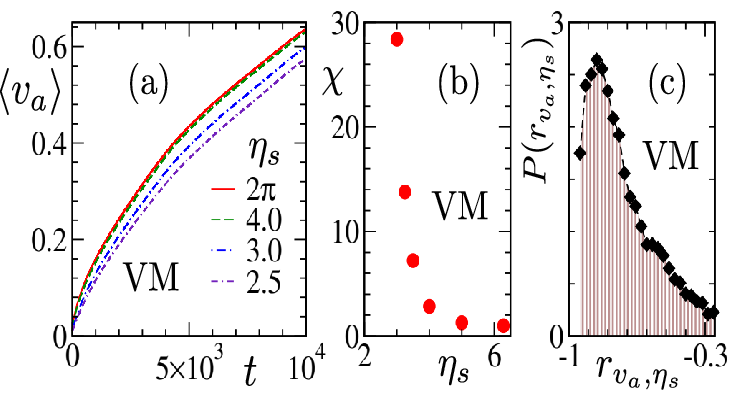}
    \caption{(a) Average Vicsek order parameter, for a set of $\eta_s$, with the variation of time. (b) Susceptibility for the VM is plotted versus $\eta_s$. (c) Distribution of the Pearson correlation coefficient concerning the evolution of $\langle{v_a}\rangle$. }
    \label{fig:velop}
\end{figure}

In the VM, the phase separation in the density field is supposed to be driven by the ordering in the velocity field.
So, it is natural to ask: Should there be faster ordering as well for higher $\eta_s$?
This indeed is the case. 
See Fig. \ref{fig:velop} (a). 
Here we plot $\langle v_a \rangle$, versus $t$, for several $\eta_s$.
The growth there implies ordering, that occurs faster for higher $\eta_s$!
The reason should again be related to fluctuation becoming critical with the approach of $\eta_s$ to $\eta_c$.
Because of voids, owing to clustering in the density field, we do not estimate correlation or domain lengths for the velocity field, rather look (see Fig. \ref{fig:velop} (b)) at the susceptibility $(\chi = (\langle v_a^2 \rangle - \langle v_a \rangle^2) L^2)$ as $\eta_s \to \eta_c$.
Clearly, there is enhancement.
Thus, for both the fields, the faster approach to new steady states are driven by the lack of critical fluctuation. 

Finally, we define a quantity $r_{v_a,\eta_s}$, analogous to $r_{t_c,\eta_s}$. 
The exercise with $r_{v_a,\eta_s}$ is simply to identify how steady the sequence of appearances of the plots is. 
For the calculated values of $r_{v_a,\eta_s}$, we have fixed $\langle v_a \rangle$ at $0.35$. 
A distribution of this is shown in Fig. \ref{fig:velop} (c). 
A message similar to Fig. \ref{fig:xi_tc} (d) clearly emerges, i.e., faster ordering for the farthest is quite general, not exhibited merely by a small fraction of initial configurations.

\section{Conclusion}

We have studied kinetics related to a class of active matter systems \cite{vicsek95} that exhibit order-disorder transitions due to aligning dynamic interaction.
As opposed to the traditional studies of coarsening phenomena \cite{bray_article, puribook}, we prepare disordered configurations at initial states having different values of the noise strength, $\eta$, the driving parameter for the transitions \cite{vicsek95}. 
Following quenches inside the ordered region, interestingly, we observe that configurations with higher values of $\eta$ reach the new steady state quicker, implying fastest arrival of the farthest.
This is counterintuitive.
Via analysis using a critical-point theory \cite{stanleybook} we show that differences in the extent of spatial correlations at initial states give rise to this puzzling effect.

It appears that certain relaxation time, associated with this effect, exhibits power-law scaling with the variation of initial correlation length.
For a comparison we have presented similar results for model para- to ferromagnetic transitions \cite{schat_potts} for which the transition is driven by the variation in temperature.
Interestingly, similar scaling exists there as well, suggesting universality.
This overall observation is analogous to the Mpemba effect \cite{mpemba} which is related to faster freezing of hotter water.
It is believed that in water and several other systems the effect appears due to reasons connecting metastability \cite{water_arxiv, bechhoffer, baity, skdLang}.
Contrary to this understanding, our work shows that metastability is not a necessary condition.

It should be noted that the enhancement in spatial correlation with approach to the transition point has counterparts in transitions in water and other materials \cite{fisher_theory, stanleybook}.
Thus, our work invites theoretical and experimental studies of coarsening dynamics in various other passive and active matters by preparing configurations at initial state points that lie along the critical loci, e.g., critical density line for water.

\section{Author Contribution}

SKD proposed the project, designed the problem, supervised the work and wrote the manuscript. SC, SD and PP wrote the codes, carried out the simulations and performed the analyses. 
TP supervised the technical aspects, alongside writing codes and producing representative results.

\section{Acknowledgement}

The authors acknowledge computation times in the clusters of National Supercomputer Mission located in JNCASR.

\bibliography{references}

\begin{thebibliography}{73}%
\makeatletter
\providecommand \@ifxundefined [1]{%
 \@ifx{#1\undefined}
}%
\providecommand \@ifnum [1]{%
 \ifnum #1\expandafter \@firstoftwo
 \else \expandafter \@secondoftwo
 \fi
}%
\providecommand \@ifx [1]{%
 \ifx #1\expandafter \@firstoftwo
 \else \expandafter \@secondoftwo
 \fi
}%
\providecommand \natexlab [1]{#1}%
\providecommand \enquote  [1]{``#1''}%
\providecommand \bibnamefont  [1]{#1}%
\providecommand \bibfnamefont [1]{#1}%
\providecommand \citenamefont [1]{#1}%
\providecommand \href@noop [0]{\@secondoftwo}%
\providecommand \href [0]{\begingroup \@sanitize@url \@href}%
\providecommand \@href[1]{\@@startlink{#1}\@@href}%
\providecommand \@@href[1]{\endgroup#1\@@endlink}%
\providecommand \@sanitize@url [0]{\catcode `\\12\catcode `\$12\catcode
  `\&12\catcode `\#12\catcode `\^12\catcode `\_12\catcode `\%12\relax}%
\providecommand \@@startlink[1]{}%
\providecommand \@@endlink[0]{}%
\providecommand \url  [0]{\begingroup\@sanitize@url \@url }%
\providecommand \@url [1]{\endgroup\@href {#1}{\urlprefix }}%
\providecommand \urlprefix  [0]{URL }%
\providecommand \Eprint [0]{\href }%
\providecommand \doibase [0]{https://doi.org/}%
\providecommand \selectlanguage [0]{\@gobble}%
\providecommand \bibinfo  [0]{\@secondoftwo}%
\providecommand \bibfield  [0]{\@secondoftwo}%
\providecommand \translation [1]{[#1]}%
\providecommand \BibitemOpen [0]{}%
\providecommand \bibitemStop [0]{}%
\providecommand \bibitemNoStop [0]{.\EOS\space}%
\providecommand \EOS [0]{\spacefactor3000\relax}%
\providecommand \BibitemShut  [1]{\csname bibitem#1\endcsname}%
\let\auto@bib@innerbib\@empty
\bibitem [{\citenamefont {Bray}(2002)}]{bray_article}%
  \BibitemOpen
  \bibfield  {author} {\bibinfo {author} {\bibfnamefont {A.~J.}\ \bibnamefont
  {Bray}},\ }\bibfield  {title} {\bibinfo {title} {{Theory of Phase-Ordering
  Kinetics}},\ }\href@noop {} {\bibfield  {journal} {\bibinfo  {journal} {Adv.
  Phys}\ }\textbf {\bibinfo {volume} {51}},\ \bibinfo {pages} {481} (\bibinfo
  {year} {2002})}\BibitemShut {NoStop}%
\bibitem [{\citenamefont {Onuki}(2002)}]{onukibook}%
  \BibitemOpen
  \bibfield  {author} {\bibinfo {author} {\bibfnamefont {A.}~\bibnamefont
  {Onuki}},\ }\href@noop {} {\emph {\bibinfo {title} {{Phase Transition
  Dynamics}}}}\ (\bibinfo  {publisher} {Cambridge University Press},\ \bibinfo
  {address} {Cambridge, UK},\ \bibinfo {year} {2002})\BibitemShut {NoStop}%
\bibitem [{\citenamefont {Puri}\ and\ \citenamefont
  {Wadhawan}(2009)}]{puribook}%
  \BibitemOpen
  \bibinfo {editor} {\bibfnamefont {S.}~\bibnamefont {Puri}}\ and\ \bibinfo
  {editor} {\bibfnamefont {V.}~\bibnamefont {Wadhawan}},\ eds.,\ \href@noop {}
  {\emph {\bibinfo {title} {{Kinetics of Phase Transitions}}}}\ (\bibinfo
  {publisher} {CRC Press},\ \bibinfo {address} {Boca Raton},\ \bibinfo {year}
  {2009})\BibitemShut {NoStop}%
\bibitem [{\citenamefont {Stenhammar}\ \emph {et~al.}(2013)\citenamefont
  {Stenhammar}, \citenamefont {Tiribocchi}, \citenamefont {Allen},
  \citenamefont {Marenduzzo},\ and\ \citenamefont {Cates}}]{Cates2013}%
  \BibitemOpen
  \bibfield  {author} {\bibinfo {author} {\bibfnamefont {J.}~\bibnamefont
  {Stenhammar}}, \bibinfo {author} {\bibfnamefont {A.}~\bibnamefont
  {Tiribocchi}}, \bibinfo {author} {\bibfnamefont {R.~J.}\ \bibnamefont
  {Allen}}, \bibinfo {author} {\bibfnamefont {D.}~\bibnamefont {Marenduzzo}},\
  and\ \bibinfo {author} {\bibfnamefont {M.~E.}\ \bibnamefont {Cates}},\
  }\bibfield  {title} {\bibinfo {title} {{Continuum Theory of Phase Separation
  Kinetics for Active Brownian Particles}},\ }\href@noop {} {\bibfield
  {journal} {\bibinfo  {journal} {Phys. Rev. Lett.}\ }\textbf {\bibinfo
  {volume} {111}},\ \bibinfo {pages} {145702} (\bibinfo {year}
  {2013})}\BibitemShut {NoStop}%
\bibitem [{\citenamefont {Redner}\ \emph {et~al.}(2013)\citenamefont {Redner},
  \citenamefont {Hagan},\ and\ \citenamefont {Baskaran}}]{Bhaskaran2013}%
  \BibitemOpen
  \bibfield  {author} {\bibinfo {author} {\bibfnamefont {G.~S.}\ \bibnamefont
  {Redner}}, \bibinfo {author} {\bibfnamefont {M.~F.}\ \bibnamefont {Hagan}},\
  and\ \bibinfo {author} {\bibfnamefont {A.}~\bibnamefont {Baskaran}},\
  }\bibfield  {title} {\bibinfo {title} {{Structure and Dynamics of a
  Phase-Separating Active Colloidal Fluid}},\ }\href@noop {} {\bibfield
  {journal} {\bibinfo  {journal} {Phys. Rev. Lett.}\ }\textbf {\bibinfo
  {volume} {110}},\ \bibinfo {pages} {055701} (\bibinfo {year}
  {2013})}\BibitemShut {NoStop}%
\bibitem [{\citenamefont {Dey}\ \emph {et~al.}(2012)\citenamefont {Dey},
  \citenamefont {Das},\ and\ \citenamefont {Rajesh}}]{Rajesh2012}%
  \BibitemOpen
  \bibfield  {author} {\bibinfo {author} {\bibfnamefont {S.}~\bibnamefont
  {Dey}}, \bibinfo {author} {\bibfnamefont {D.}~\bibnamefont {Das}},\ and\
  \bibinfo {author} {\bibfnamefont {R.}~\bibnamefont {Rajesh}},\ }\bibfield
  {title} {\bibinfo {title} {{Spatial Structures and Giant Number Fluctuations
  in Models of Active Matter}},\ }\href@noop {} {\bibfield  {journal} {\bibinfo
   {journal} {Phys. Rev. Lett.}\ }\textbf {\bibinfo {volume} {108}},\ \bibinfo
  {pages} {238001} (\bibinfo {year} {2012})}\BibitemShut {NoStop}%
\bibitem [{\citenamefont {Das}\ \emph {et~al.}(2014)\citenamefont {Das},
  \citenamefont {Egorov}, \citenamefont {Trefz}, \citenamefont {Virnau},\ and\
  \citenamefont {Binder}}]{das14}%
  \BibitemOpen
  \bibfield  {author} {\bibinfo {author} {\bibfnamefont {S.~K.}\ \bibnamefont
  {Das}}, \bibinfo {author} {\bibfnamefont {S.~A.}\ \bibnamefont {Egorov}},
  \bibinfo {author} {\bibfnamefont {B.}~\bibnamefont {Trefz}}, \bibinfo
  {author} {\bibfnamefont {P.}~\bibnamefont {Virnau}},\ and\ \bibinfo {author}
  {\bibfnamefont {K.}~\bibnamefont {Binder}},\ }\bibfield  {title} {\bibinfo
  {title} {{Phase Behavior of Active Swimmers in Depletants: {M}olecular
  Dynamics and Integral Equation Theory}},\ }\href@noop {} {\bibfield
  {journal} {\bibinfo  {journal} {Phys. Rev. Lett.}\ }\textbf {\bibinfo
  {volume} {112}},\ \bibinfo {pages} {198301} (\bibinfo {year}
  {2014})}\BibitemShut {NoStop}%
\bibitem [{\citenamefont {Das}(2017)}]{das2017}%
  \BibitemOpen
  \bibfield  {author} {\bibinfo {author} {\bibfnamefont {S.~K.}\ \bibnamefont
  {Das}},\ }\bibfield  {title} {\bibinfo {title} {{Pattern, Growth, and Aging
  in Aggregation Kinetics of a Vicsek-like Active Matter Model}},\ }\href@noop
  {} {\bibfield  {journal} {\bibinfo  {journal} {J. Chem. Phys.}\ }\textbf
  {\bibinfo {volume} {146}},\ \bibinfo {pages} {044902} (\bibinfo {year}
  {2017})}\BibitemShut {NoStop}%
\bibitem [{\citenamefont {Paul}\ \emph {et~al.}(2021)\citenamefont {Paul},
  \citenamefont {Bera},\ and\ \citenamefont {Das}}]{paul_bera}%
  \BibitemOpen
  \bibfield  {author} {\bibinfo {author} {\bibfnamefont {S.}~\bibnamefont
  {Paul}}, \bibinfo {author} {\bibfnamefont {A.}~\bibnamefont {Bera}},\ and\
  \bibinfo {author} {\bibfnamefont {S.~K.}\ \bibnamefont {Das}},\ }\bibfield
  {title} {\bibinfo {title} {{How do clusters in Phase-Separating Active Matter
  Systems Grow? A Study for Vicsek Activity in Systems Undergoing Vapor--Solid
  Transition}},\ }\href@noop {} {\bibfield  {journal} {\bibinfo  {journal}
  {Soft Matter}\ }\textbf {\bibinfo {volume} {17}},\ \bibinfo {pages} {645}
  (\bibinfo {year} {2021})}\BibitemShut {NoStop}%
\bibitem [{\citenamefont {Katyal}\ \emph {et~al.}(2020)\citenamefont {Katyal},
  \citenamefont {Dey}, \citenamefont {Das},\ and\ \citenamefont
  {Puri}}]{puricoarsening}%
  \BibitemOpen
  \bibfield  {author} {\bibinfo {author} {\bibfnamefont {N.}~\bibnamefont
  {Katyal}}, \bibinfo {author} {\bibfnamefont {S.}~\bibnamefont {Dey}},
  \bibinfo {author} {\bibfnamefont {D.}~\bibnamefont {Das}},\ and\ \bibinfo
  {author} {\bibfnamefont {S.}~\bibnamefont {Puri}},\ }\bibfield  {title}
  {\bibinfo {title} {{Coarsening Dynamics in the Vicsek Model of Active
  Matter}},\ }\href@noop {} {\bibfield  {journal} {\bibinfo  {journal} {Eur.
  Phys. J. E}\ }\textbf {\bibinfo {volume} {43}},\ \bibinfo {pages} {10}
  (\bibinfo {year} {2020})}\BibitemShut {NoStop}%
\bibitem [{\citenamefont {Paul}\ \emph {et~al.}(2024)\citenamefont {Paul},
  \citenamefont {Vadakkayil},\ and\ \citenamefont {Das}}]{TP}%
  \BibitemOpen
  \bibfield  {author} {\bibinfo {author} {\bibfnamefont {T.}~\bibnamefont
  {Paul}}, \bibinfo {author} {\bibfnamefont {N.}~\bibnamefont {Vadakkayil}},\
  and\ \bibinfo {author} {\bibfnamefont {S.~K.}\ \bibnamefont {Das}},\
  }\bibfield  {title} {\bibinfo {title} {{Finite-Size Scaling in Kinetics of
  Phase Separation in Certain Models of Aligning Active Particles}},\
  }\href@noop {} {\bibfield  {journal} {\bibinfo  {journal} {Phys. Rev. E}\
  }\textbf {\bibinfo {volume} {109}},\ \bibinfo {pages} {064607} (\bibinfo
  {year} {2024})}\BibitemShut {NoStop}%
\bibitem [{\citenamefont {Cremer}\ and\ \citenamefont
  {L\"owen}(2014)}]{lowen_pre}%
  \BibitemOpen
  \bibfield  {author} {\bibinfo {author} {\bibfnamefont {P.}~\bibnamefont
  {Cremer}}\ and\ \bibinfo {author} {\bibfnamefont {H.}~\bibnamefont
  {L\"owen}},\ }\bibfield  {title} {\bibinfo {title} {{Scaling of Cluster
  Growth for Coagulating Active Particles}},\ }\href@noop {} {\bibfield
  {journal} {\bibinfo  {journal} {Phys. Rev. E}\ }\textbf {\bibinfo {volume}
  {89}},\ \bibinfo {pages} {022307} (\bibinfo {year} {2014})}\BibitemShut
  {NoStop}%
\bibitem [{\citenamefont {Peruani}\ and\ \citenamefont
  {Baer}(2013)}]{activekinetic}%
  \BibitemOpen
  \bibfield  {author} {\bibinfo {author} {\bibfnamefont {F.}~\bibnamefont
  {Peruani}}\ and\ \bibinfo {author} {\bibfnamefont {M.}~\bibnamefont {Baer}},\
  }\bibfield  {title} {\bibinfo {title} {{A Kinetic Model and Scaling
  Properties of Non-Equilibrium Clustering of Self-Propelled Particles}},\
  }\href@noop {} {\bibfield  {journal} {\bibinfo  {journal} {New J. Phys.}\
  }\textbf {\bibinfo {volume} {15}},\ \bibinfo {pages} {065009} (\bibinfo
  {year} {2013})}\BibitemShut {NoStop}%
\bibitem [{\citenamefont {Surrey}\ \emph {et~al.}(2001)\citenamefont {Surrey},
  \citenamefont {N{\'e}d{\'e}lec}, \citenamefont {Leibler},\ and\ \citenamefont
  {Karsenti}}]{experiment0}%
  \BibitemOpen
  \bibfield  {author} {\bibinfo {author} {\bibfnamefont {T.}~\bibnamefont
  {Surrey}}, \bibinfo {author} {\bibfnamefont {F.}~\bibnamefont
  {N{\'e}d{\'e}lec}}, \bibinfo {author} {\bibfnamefont {S.}~\bibnamefont
  {Leibler}},\ and\ \bibinfo {author} {\bibfnamefont {E.}~\bibnamefont
  {Karsenti}},\ }\bibfield  {title} {\bibinfo {title} {{Physical Properties
  Determining Self-Organization of Motors and Microtubules}},\ }\href@noop {}
  {\bibfield  {journal} {\bibinfo  {journal} {Science}\ }\textbf {\bibinfo
  {volume} {292}},\ \bibinfo {pages} {1167} (\bibinfo {year}
  {2001})}\BibitemShut {NoStop}%
\bibitem [{\citenamefont {Buttinoni}\ \emph {et~al.}(2013)\citenamefont
  {Buttinoni}, \citenamefont {Bialk{\'e}}, \citenamefont {K{\"u}mmel},
  \citenamefont {L{\"o}wen}, \citenamefont {Bechinger},\ and\ \citenamefont
  {Speck}}]{experiment1}%
  \BibitemOpen
  \bibfield  {author} {\bibinfo {author} {\bibfnamefont {I.}~\bibnamefont
  {Buttinoni}}, \bibinfo {author} {\bibfnamefont {J.}~\bibnamefont
  {Bialk{\'e}}}, \bibinfo {author} {\bibfnamefont {F.}~\bibnamefont
  {K{\"u}mmel}}, \bibinfo {author} {\bibfnamefont {H.}~\bibnamefont
  {L{\"o}wen}}, \bibinfo {author} {\bibfnamefont {C.}~\bibnamefont
  {Bechinger}},\ and\ \bibinfo {author} {\bibfnamefont {T.}~\bibnamefont
  {Speck}},\ }\bibfield  {title} {\bibinfo {title} {{Dynamical Clustering and
  Phase Separation in Suspensions of Self-Propelled Colloidal Particles}},\
  }\href@noop {} {\bibfield  {journal} {\bibinfo  {journal} {Phys. Rev. Lett.}\
  }\textbf {\bibinfo {volume} {110}},\ \bibinfo {pages} {238301} (\bibinfo
  {year} {2013})}\BibitemShut {NoStop}%
\bibitem [{\citenamefont {Huber}\ \emph {et~al.}(2018)\citenamefont {Huber},
  \citenamefont {Suzuki}, \citenamefont {Kr{\"u}ger}, \citenamefont {Frey},\
  and\ \citenamefont {Bausch}}]{experiment2}%
  \BibitemOpen
  \bibfield  {author} {\bibinfo {author} {\bibfnamefont {L.}~\bibnamefont
  {Huber}}, \bibinfo {author} {\bibfnamefont {R.}~\bibnamefont {Suzuki}},
  \bibinfo {author} {\bibfnamefont {T.}~\bibnamefont {Kr{\"u}ger}}, \bibinfo
  {author} {\bibfnamefont {E.}~\bibnamefont {Frey}},\ and\ \bibinfo {author}
  {\bibfnamefont {A.}~\bibnamefont {Bausch}},\ }\bibfield  {title} {\bibinfo
  {title} {{Emergence of Coexisting Ordered States in Active Matter Systems}},\
  }\href@noop {} {\bibfield  {journal} {\bibinfo  {journal} {Science}\ }\textbf
  {\bibinfo {volume} {361}},\ \bibinfo {pages} {255} (\bibinfo {year}
  {2018})}\BibitemShut {NoStop}%
\bibitem [{\citenamefont {Grauer}\ \emph {et~al.}(2021)\citenamefont {Grauer},
  \citenamefont {Schmidt}, \citenamefont {Pineda}, \citenamefont {Midtvedt},
  \citenamefont {L{\"o}wen}, \citenamefont {Volpe},\ and\ \citenamefont
  {Liebchen}}]{experiment3}%
  \BibitemOpen
  \bibfield  {author} {\bibinfo {author} {\bibfnamefont {J.}~\bibnamefont
  {Grauer}}, \bibinfo {author} {\bibfnamefont {F.}~\bibnamefont {Schmidt}},
  \bibinfo {author} {\bibfnamefont {J.}~\bibnamefont {Pineda}}, \bibinfo
  {author} {\bibfnamefont {B.}~\bibnamefont {Midtvedt}}, \bibinfo {author}
  {\bibfnamefont {H.}~\bibnamefont {L{\"o}wen}}, \bibinfo {author}
  {\bibfnamefont {G.}~\bibnamefont {Volpe}},\ and\ \bibinfo {author}
  {\bibfnamefont {B.}~\bibnamefont {Liebchen}},\ }\bibfield  {title} {\bibinfo
  {title} {{Active Droploids}},\ }\href@noop {} {\bibfield  {journal} {\bibinfo
   {journal} {Nat. Commun.}\ }\textbf {\bibinfo {volume} {12}},\ \bibinfo
  {pages} {6005} (\bibinfo {year} {2021})}\BibitemShut {NoStop}%
\bibitem [{\citenamefont {Bera}\ \emph {et~al.}(2022)\citenamefont {Bera},
  \citenamefont {Sahoo}, \citenamefont {Thakur},\ and\ \citenamefont
  {Das}}]{bera}%
  \BibitemOpen
  \bibfield  {author} {\bibinfo {author} {\bibfnamefont {A.}~\bibnamefont
  {Bera}}, \bibinfo {author} {\bibfnamefont {S.}~\bibnamefont {Sahoo}},
  \bibinfo {author} {\bibfnamefont {S.}~\bibnamefont {Thakur}},\ and\ \bibinfo
  {author} {\bibfnamefont {S.~K.}\ \bibnamefont {Das}},\ }\bibfield  {title}
  {\bibinfo {title} {{Active Particles in Explicit Solvent: Dynamics of
  Clustering for Alignment Interaction}},\ }\href@noop {} {\bibfield  {journal}
  {\bibinfo  {journal} {Phys. Rev. E}\ }\textbf {\bibinfo {volume} {105}},\
  \bibinfo {pages} {014606} (\bibinfo {year} {2022})}\BibitemShut {NoStop}%
\bibitem [{\citenamefont {Mishra}\ \emph {et~al.}(2014)\citenamefont {Mishra},
  \citenamefont {Puri},\ and\ \citenamefont {Ramaswamy}}]{mishra}%
  \BibitemOpen
  \bibfield  {author} {\bibinfo {author} {\bibfnamefont {S.}~\bibnamefont
  {Mishra}}, \bibinfo {author} {\bibfnamefont {S.}~\bibnamefont {Puri}},\ and\
  \bibinfo {author} {\bibfnamefont {S.}~\bibnamefont {Ramaswamy}},\ }\bibfield
  {title} {\bibinfo {title} {{Aspects of the Density Field in an Active
  Nematic}},\ }\href@noop {} {\bibfield  {journal} {\bibinfo  {journal}
  {Philos. Trans. A: Math. Phys. Eng. Sci.}\ }\textbf {\bibinfo {volume}
  {372}},\ \bibinfo {pages} {20130364} (\bibinfo {year} {2014})}\BibitemShut
  {NoStop}%
\bibitem [{\citenamefont {Parameshwaran}\ and\ \citenamefont
  {Gupta}(2024)}]{bhaskar}%
  \BibitemOpen
  \bibfield  {author} {\bibinfo {author} {\bibfnamefont {A.}~\bibnamefont
  {Parameshwaran}}\ and\ \bibinfo {author} {\bibfnamefont {B.~S.}\ \bibnamefont
  {Gupta}},\ }\bibfield  {title} {\bibinfo {title} {{Kinetics of Vapor-Liquid
  Transition of Active Matter System Under Quasi One-Dimensional
  Confinement}},\ }\href@noop {} {\bibfield  {journal} {\bibinfo  {journal}
  {arXiv:2408.01195}\ } (\bibinfo {year} {2024})}\BibitemShut {NoStop}%
\bibitem [{\citenamefont {Solon}\ \emph {et~al.}(2015)\citenamefont {Solon},
  \citenamefont {Chat\'e},\ and\ \citenamefont {Tailleur}}]{chate}%
  \BibitemOpen
  \bibfield  {author} {\bibinfo {author} {\bibfnamefont {A.~P.}\ \bibnamefont
  {Solon}}, \bibinfo {author} {\bibfnamefont {H.}~\bibnamefont {Chat\'e}},\
  and\ \bibinfo {author} {\bibfnamefont {J.}~\bibnamefont {Tailleur}},\
  }\bibfield  {title} {\bibinfo {title} {{From Phase to Microphase Separation
  in Flocking Models: The Essential Role of Nonequilibrium Fluctuations}},\
  }\href@noop {} {\bibfield  {journal} {\bibinfo  {journal} {Phys. Rev. Lett.}\
  }\textbf {\bibinfo {volume} {114}},\ \bibinfo {pages} {068101} (\bibinfo
  {year} {2015})}\BibitemShut {NoStop}%
\bibitem [{\citenamefont {Paul}\ \emph {et~al.}(2022)\citenamefont {Paul},
  \citenamefont {Majumder}, \citenamefont {Das},\ and\ \citenamefont
  {Janke}}]{paul2022}%
  \BibitemOpen
  \bibfield  {author} {\bibinfo {author} {\bibfnamefont {S.}~\bibnamefont
  {Paul}}, \bibinfo {author} {\bibfnamefont {S.}~\bibnamefont {Majumder}},
  \bibinfo {author} {\bibfnamefont {S.~K.}\ \bibnamefont {Das}},\ and\ \bibinfo
  {author} {\bibfnamefont {W.}~\bibnamefont {Janke}},\ }\bibfield  {title}
  {\bibinfo {title} {{Effects of Alignment Activity on the Collapse Kinetics of
  a Flexible Polymer}},\ }\href@noop {} {\bibfield  {journal} {\bibinfo
  {journal} {Soft Matter}\ }\textbf {\bibinfo {volume} {18}},\ \bibinfo {pages}
  {1978} (\bibinfo {year} {2022})}\BibitemShut {NoStop}%
\bibitem [{\citenamefont {Caporusso}\ \emph {et~al.}(2023)\citenamefont
  {Caporusso}, \citenamefont {Cugliandolo}, \citenamefont {Digregorio},
  \citenamefont {Gonnella}, \citenamefont {Levis},\ and\ \citenamefont
  {Suma}}]{miclusters}%
  \BibitemOpen
  \bibfield  {author} {\bibinfo {author} {\bibfnamefont {C.~B.}\ \bibnamefont
  {Caporusso}}, \bibinfo {author} {\bibfnamefont {L.~F.}\ \bibnamefont
  {Cugliandolo}}, \bibinfo {author} {\bibfnamefont {P.}~\bibnamefont
  {Digregorio}}, \bibinfo {author} {\bibfnamefont {G.}~\bibnamefont
  {Gonnella}}, \bibinfo {author} {\bibfnamefont {D.}~\bibnamefont {Levis}},\
  and\ \bibinfo {author} {\bibfnamefont {A.}~\bibnamefont {Suma}},\ }\bibfield
  {title} {\bibinfo {title} {{Dynamics of Motility-Induced Clusters: Coarsening
  beyond Ostwald Ripening}},\ }\href@noop {} {\bibfield  {journal} {\bibinfo
  {journal} {Phys. Rev. Lett.}\ }\textbf {\bibinfo {volume} {131}},\ \bibinfo
  {pages} {068201} (\bibinfo {year} {2023})}\BibitemShut {NoStop}%
\bibitem [{\citenamefont {Fisher}(1967)}]{fisher_theory}%
  \BibitemOpen
  \bibfield  {author} {\bibinfo {author} {\bibfnamefont {M.~E.}\ \bibnamefont
  {Fisher}},\ }\bibfield  {title} {\bibinfo {title} {{The Theory of Equilibrium
  Critical Phenomena}},\ }\href@noop {} {\bibfield  {journal} {\bibinfo
  {journal} {Rep. Prog. Phys.}\ }\textbf {\bibinfo {volume} {30}},\ \bibinfo
  {pages} {615} (\bibinfo {year} {1967})}\BibitemShut {NoStop}%
\bibitem [{\citenamefont {Bray}\ \emph {et~al.}(1991)\citenamefont {Bray},
  \citenamefont {Humayun},\ and\ \citenamefont {Newman}}]{Bray_Tc}%
  \BibitemOpen
  \bibfield  {author} {\bibinfo {author} {\bibfnamefont {A.~J.}\ \bibnamefont
  {Bray}}, \bibinfo {author} {\bibfnamefont {K.}~\bibnamefont {Humayun}},\ and\
  \bibinfo {author} {\bibfnamefont {T.~J.}\ \bibnamefont {Newman}},\ }\bibfield
   {title} {\bibinfo {title} {{Kinetics of Ordering for Correlated Initial
  Conditions}},\ }\href@noop {} {\bibfield  {journal} {\bibinfo  {journal}
  {Phys. Rev. B}\ }\textbf {\bibinfo {volume} {43}},\ \bibinfo {pages} {3699}
  (\bibinfo {year} {1991})}\BibitemShut {NoStop}%
\bibitem [{\citenamefont {Blanchard}\ \emph {et~al.}(2014)\citenamefont
  {Blanchard}, \citenamefont {Cugliandolo},\ and\ \citenamefont
  {Picco}}]{Cugliandolo_Tc}%
  \BibitemOpen
  \bibfield  {author} {\bibinfo {author} {\bibfnamefont {T.}~\bibnamefont
  {Blanchard}}, \bibinfo {author} {\bibfnamefont {L.~F.}\ \bibnamefont
  {Cugliandolo}},\ and\ \bibinfo {author} {\bibfnamefont {M.}~\bibnamefont
  {Picco}},\ }\bibfield  {title} {\bibinfo {title} {{Persistence in the Two
  Dimensional Ferromagnetic Ising Model}},\ }\href@noop {} {\bibfield
  {journal} {\bibinfo  {journal} {J. Stat. Mech: Theor. Expt.}\ }\textbf
  {\bibinfo {volume} {2014}},\ \bibinfo {pages} {P12021} (\bibinfo {year}
  {2014})}\BibitemShut {NoStop}%
\bibitem [{\citenamefont {Chakraborty}\ and\ \citenamefont
  {Das}(2015)}]{saikat_EPJB}%
  \BibitemOpen
  \bibfield  {author} {\bibinfo {author} {\bibfnamefont {S.}~\bibnamefont
  {Chakraborty}}\ and\ \bibinfo {author} {\bibfnamefont {S.~K.}\ \bibnamefont
  {Das}},\ }\bibfield  {title} {\bibinfo {title} {{Role of Initial Correlation
  in Coarsening of a Ferromagnet}},\ }\href@noop {} {\bibfield  {journal}
  {\bibinfo  {journal} {Eur. Phys. J. B}\ }\textbf {\bibinfo {volume} {88}},\
  \bibinfo {pages} {160} (\bibinfo {year} {2015})}\BibitemShut {NoStop}%
\bibitem [{\citenamefont {Das}\ \emph {et~al.}(2020{\natexlab{a}})\citenamefont
  {Das}, \citenamefont {Das}, \citenamefont {Vadakkayil}, \citenamefont
  {Chakraborty},\ and\ \citenamefont {Paul}}]{Das_JPCM}%
  \BibitemOpen
  \bibfield  {author} {\bibinfo {author} {\bibfnamefont {S.~K.}\ \bibnamefont
  {Das}}, \bibinfo {author} {\bibfnamefont {K.}~\bibnamefont {Das}}, \bibinfo
  {author} {\bibfnamefont {N.}~\bibnamefont {Vadakkayil}}, \bibinfo {author}
  {\bibfnamefont {S.}~\bibnamefont {Chakraborty}},\ and\ \bibinfo {author}
  {\bibfnamefont {S.}~\bibnamefont {Paul}},\ }\bibfield  {title} {\bibinfo
  {title} {{Initial Correlation Dependence of Aging in Phase Separating Solid
  Binary Mixtures and Ordering Ferromagnets}},\ }\href@noop {} {\bibfield
  {journal} {\bibinfo  {journal} {J. Phys. Condens. Matter}\ }\textbf {\bibinfo
  {volume} {32}},\ \bibinfo {pages} {184005} (\bibinfo {year}
  {2020}{\natexlab{a}})}\BibitemShut {NoStop}%
\bibitem [{\citenamefont {Das}\ \emph {et~al.}(2020{\natexlab{b}})\citenamefont
  {Das}, \citenamefont {Vadakkayil},\ and\ \citenamefont {Das}}]{koyelina_PRE}%
  \BibitemOpen
  \bibfield  {author} {\bibinfo {author} {\bibfnamefont {K.}~\bibnamefont
  {Das}}, \bibinfo {author} {\bibfnamefont {N.}~\bibnamefont {Vadakkayil}},\
  and\ \bibinfo {author} {\bibfnamefont {S.~K.}\ \bibnamefont {Das}},\
  }\bibfield  {title} {\bibinfo {title} {{Aging Exponents for Nonequilibrium
  Dynamics Following Quenches from Critical Points}},\ }\href@noop {}
  {\bibfield  {journal} {\bibinfo  {journal} {Phys. Rev. E}\ }\textbf {\bibinfo
  {volume} {101}},\ \bibinfo {pages} {062112} (\bibinfo {year}
  {2020}{\natexlab{b}})}\BibitemShut {NoStop}%
\bibitem [{\citenamefont {Das}(2023)}]{skdLang}%
  \BibitemOpen
  \bibfield  {author} {\bibinfo {author} {\bibfnamefont {S.~K.}\ \bibnamefont
  {Das}},\ }\bibfield  {title} {\bibinfo {title} {{Perspectives on Certain
  Puzzles in Phase Transformations: {When} Should the Farthest Reach the
  Earliest?}},\ }\href@noop {} {\bibfield  {journal} {\bibinfo  {journal}
  {Langmuir}\ }\textbf {\bibinfo {volume} {39}},\ \bibinfo {pages} {10715}
  (\bibinfo {year} {2023})}\BibitemShut {NoStop}%
\bibitem [{\citenamefont {Vicsek}\ \emph {et~al.}(1995)\citenamefont {Vicsek},
  \citenamefont {Czir\'ok}, \citenamefont {Ben-Jacob}, \citenamefont {Cohen},\
  and\ \citenamefont {Shochet}}]{vicsek95}%
  \BibitemOpen
  \bibfield  {author} {\bibinfo {author} {\bibfnamefont {T.}~\bibnamefont
  {Vicsek}}, \bibinfo {author} {\bibfnamefont {A.}~\bibnamefont {Czir\'ok}},
  \bibinfo {author} {\bibfnamefont {E.}~\bibnamefont {Ben-Jacob}}, \bibinfo
  {author} {\bibfnamefont {I.}~\bibnamefont {Cohen}},\ and\ \bibinfo {author}
  {\bibfnamefont {O.}~\bibnamefont {Shochet}},\ }\bibfield  {title} {\bibinfo
  {title} {{Novel Type of Phase Transition in a System of Self-Driven
  Particles}},\ }\href@noop {} {\bibfield  {journal} {\bibinfo  {journal}
  {Phys. Rev. Lett.}\ }\textbf {\bibinfo {volume} {75}},\ \bibinfo {pages}
  {1226} (\bibinfo {year} {1995})}\BibitemShut {NoStop}%
\bibitem [{\citenamefont {Czir{\'o}k}\ \emph {et~al.}(1997)\citenamefont
  {Czir{\'o}k}, \citenamefont {Stanley},\ and\ \citenamefont
  {Vicsek}}]{vicsek1997}%
  \BibitemOpen
  \bibfield  {author} {\bibinfo {author} {\bibfnamefont {A.}~\bibnamefont
  {Czir{\'o}k}}, \bibinfo {author} {\bibfnamefont {H.~E.}\ \bibnamefont
  {Stanley}},\ and\ \bibinfo {author} {\bibfnamefont {T.}~\bibnamefont
  {Vicsek}},\ }\bibfield  {title} {\bibinfo {title} {{Spontaneously Ordered
  Motion of Self-Propelled Particles}},\ }\href@noop {} {\bibfield  {journal}
  {\bibinfo  {journal} {J. Phys. A Math. Gen.}\ }\textbf {\bibinfo {volume}
  {30}},\ \bibinfo {pages} {1375} (\bibinfo {year} {1997})}\BibitemShut
  {NoStop}%
\bibitem [{\citenamefont {Vicsek}\ and\ \citenamefont
  {Zafeiris}(2012)}]{Vicsek_rep2012}%
  \BibitemOpen
  \bibfield  {author} {\bibinfo {author} {\bibfnamefont {T.}~\bibnamefont
  {Vicsek}}\ and\ \bibinfo {author} {\bibfnamefont {A.}~\bibnamefont
  {Zafeiris}},\ }\bibfield  {title} {\bibinfo {title} {{Collective Motion}},\
  }\href@noop {} {\bibfield  {journal} {\bibinfo  {journal} {Phys. Rep.}\
  }\textbf {\bibinfo {volume} {517}},\ \bibinfo {pages} {71} (\bibinfo {year}
  {2012})}\BibitemShut {NoStop}%
\bibitem [{\citenamefont {Stanley}(1971)}]{stanleybook}%
  \BibitemOpen
  \bibfield  {author} {\bibinfo {author} {\bibfnamefont {H.~E.}\ \bibnamefont
  {Stanley}},\ }\href@noop {} {\emph {\bibinfo {title} {{Introduction to Phase
  Transitions and Critical Phenomena}}}}\ (\bibinfo  {publisher} {Clarendon
  Press},\ \bibinfo {address} {Oxford},\ \bibinfo {year} {1971})\BibitemShut
  {NoStop}%
\bibitem [{\citenamefont {Mpemba}\ and\ \citenamefont
  {Osborne}(1969)}]{mpemba}%
  \BibitemOpen
  \bibfield  {author} {\bibinfo {author} {\bibfnamefont {E.~B.}\ \bibnamefont
  {Mpemba}}\ and\ \bibinfo {author} {\bibfnamefont {D.~G.}\ \bibnamefont
  {Osborne}},\ }\bibfield  {title} {\bibinfo {title} {{Cool?}},\ }\href@noop {}
  {\bibfield  {journal} {\bibinfo  {journal} {Physics Education}\ }\textbf
  {\bibinfo {volume} {4}},\ \bibinfo {pages} {172} (\bibinfo {year}
  {1969})}\BibitemShut {NoStop}%
\bibitem [{\citenamefont {Aristotle}(1962)}]{aristotle}%
  \BibitemOpen
  \bibfield  {author} {\bibinfo {author} {\bibnamefont {Aristotle}},\
  }\href@noop {} {\emph {\bibinfo {title} {{Meteorologica}}}},\ edited by\
  \bibinfo {editor} {\bibfnamefont {H.~D.}\ \bibnamefont {Lee}}\ (\bibinfo
  {publisher} {Harvard University Press},\ \bibinfo {address} {Cambridge},\
  \bibinfo {year} {1962})\BibitemShut {NoStop}%
\bibitem [{\citenamefont {Ghosh}\ \emph {et~al.}(2024)\citenamefont {Ghosh},
  \citenamefont {Pathak}, \citenamefont {Chatterjee},\ and\ \citenamefont
  {Das}}]{water_arxiv}%
  \BibitemOpen
  \bibfield  {author} {\bibinfo {author} {\bibfnamefont {S.}~\bibnamefont
  {Ghosh}}, \bibinfo {author} {\bibfnamefont {P.}~\bibnamefont {Pathak}},
  \bibinfo {author} {\bibfnamefont {S.}~\bibnamefont {Chatterjee}},\ and\
  \bibinfo {author} {\bibfnamefont {S.~K.}\ \bibnamefont {Das}},\ }\bibfield
  {title} {\bibinfo {title} {{Simulations of {Mpemba} Effect in {WATER},
  {Lennard-Jones and Ising} Models: {Metastability} vs Critical
  Fluctuations}},\ }\href@noop {} {\bibfield  {journal} {\bibinfo  {journal}
  {arXiv:2407.06954v1}\ } (\bibinfo {year} {2024})}\BibitemShut {NoStop}%
\bibitem [{\citenamefont {Bechhoefer}\ \emph {et~al.}(2021)\citenamefont
  {Bechhoefer}, \citenamefont {Kumar},\ and\ \citenamefont
  {Ch{\'e}trite}}]{bechhoffer}%
  \BibitemOpen
  \bibfield  {author} {\bibinfo {author} {\bibfnamefont {J.}~\bibnamefont
  {Bechhoefer}}, \bibinfo {author} {\bibfnamefont {A.}~\bibnamefont {Kumar}},\
  and\ \bibinfo {author} {\bibfnamefont {R.}~\bibnamefont {Ch{\'e}trite}},\
  }\bibfield  {title} {\bibinfo {title} {{A Fresh Understanding of the {Mpemba}
  Effect}},\ }\href@noop {} {\bibfield  {journal} {\bibinfo  {journal} {Nat.
  Rev. Phys.}\ }\textbf {\bibinfo {volume} {3}},\ \bibinfo {pages} {534}
  (\bibinfo {year} {2021})}\BibitemShut {NoStop}%
\bibitem [{\citenamefont {Jeng}(2006)}]{jeng}%
  \BibitemOpen
  \bibfield  {author} {\bibinfo {author} {\bibfnamefont {M.}~\bibnamefont
  {Jeng}},\ }\bibfield  {title} {\bibinfo {title} {{The {Mpemba} Effect: {When}
  Can Hot Water Freeze Faster Than Cold?}},\ }\href@noop {} {\bibfield
  {journal} {\bibinfo  {journal} {Am. J. Phys.}\ }\textbf {\bibinfo {volume}
  {74}},\ \bibinfo {pages} {514} (\bibinfo {year} {2006})}\BibitemShut
  {NoStop}%
\bibitem [{\citenamefont {{M. {Baity-Jesi} et. al.}}(2019)}]{baity}%
  \BibitemOpen
  \bibfield  {author} {\bibinfo {author} {\bibnamefont {{M. {Baity-Jesi} et.
  al.}}},\ }\bibfield  {title} {\bibinfo {title} {{The Mpemba Effect in Spin
  Glasses is a Persistent Memory Effect}},\ }\href@noop {} {\bibfield
  {journal} {\bibinfo  {journal} {Proc. Natl. Acad. Sci. U. S. A.}\ }\textbf
  {\bibinfo {volume} {116}},\ \bibinfo {pages} {15350} (\bibinfo {year}
  {2019})}\BibitemShut {NoStop}%
\bibitem [{\citenamefont {Vadakkayil}\ and\ \citenamefont {Das}(2021)}]{nv}%
  \BibitemOpen
  \bibfield  {author} {\bibinfo {author} {\bibfnamefont {N.}~\bibnamefont
  {Vadakkayil}}\ and\ \bibinfo {author} {\bibfnamefont {S.~K.}\ \bibnamefont
  {Das}},\ }\bibfield  {title} {\bibinfo {title} {{Should a Hotter Paramagnet
  Transform Quicker To a Ferromagnet? {Monte} Carlo Simulation Results for
  {Ising} Model}},\ }\href@noop {} {\bibfield  {journal} {\bibinfo  {journal}
  {Phys. Chem. Chem. Phys.}\ }\textbf {\bibinfo {volume} {23}},\ \bibinfo
  {pages} {11186} (\bibinfo {year} {2021})}\BibitemShut {NoStop}%
\bibitem [{\citenamefont {Chatterjee}\ \emph {et~al.}(2024)\citenamefont
  {Chatterjee}, \citenamefont {Ghosh}, \citenamefont {Vadakkayil},
  \citenamefont {Paul}, \citenamefont {Singha},\ and\ \citenamefont
  {Das}}]{schat_potts}%
  \BibitemOpen
  \bibfield  {author} {\bibinfo {author} {\bibfnamefont {S.}~\bibnamefont
  {Chatterjee}}, \bibinfo {author} {\bibfnamefont {S.}~\bibnamefont {Ghosh}},
  \bibinfo {author} {\bibfnamefont {N.}~\bibnamefont {Vadakkayil}}, \bibinfo
  {author} {\bibfnamefont {T.}~\bibnamefont {Paul}}, \bibinfo {author}
  {\bibfnamefont {S.~K.}\ \bibnamefont {Singha}},\ and\ \bibinfo {author}
  {\bibfnamefont {S.~K.}\ \bibnamefont {Das}},\ }\bibfield  {title} {\bibinfo
  {title} {{Mpemba effect in Pure Spin Systems : A Universal Picture of the
  Role of Spatial Correlations at Initial States}},\ }\href@noop {} {\bibfield
  {journal} {\bibinfo  {journal} {Phys. Rev. E}\ }\textbf {\bibinfo {volume}
  {110}},\ \bibinfo {pages} {L012103} (\bibinfo {year} {2024})}\BibitemShut
  {NoStop}%
\bibitem [{\citenamefont {Gal}\ and\ \citenamefont {Raz}(2020)}]{gal_raz}%
  \BibitemOpen
  \bibfield  {author} {\bibinfo {author} {\bibfnamefont {A.}~\bibnamefont
  {Gal}}\ and\ \bibinfo {author} {\bibfnamefont {O.}~\bibnamefont {Raz}},\
  }\bibfield  {title} {\bibinfo {title} {{Precooling Strategy Allows
  Exponentially Faster Heating}},\ }\href@noop {} {\bibfield  {journal}
  {\bibinfo  {journal} {Phys. Rev. Lett.}\ }\textbf {\bibinfo {volume} {124}},\
  \bibinfo {pages} {060602} (\bibinfo {year} {2020})}\BibitemShut {NoStop}%
\bibitem [{\citenamefont {Chaddah}\ \emph {et~al.}(2010)\citenamefont
  {Chaddah}, \citenamefont {Dash}, \citenamefont {Kumar},\ and\ \citenamefont
  {Banerjee}}]{chaddah}%
  \BibitemOpen
  \bibfield  {author} {\bibinfo {author} {\bibfnamefont {P.}~\bibnamefont
  {Chaddah}}, \bibinfo {author} {\bibfnamefont {S.}~\bibnamefont {Dash}},
  \bibinfo {author} {\bibfnamefont {K.}~\bibnamefont {Kumar}},\ and\ \bibinfo
  {author} {\bibfnamefont {A.}~\bibnamefont {Banerjee}},\ }\bibfield  {title}
  {\bibinfo {title} {{Overtaking while Approaching Equilibrium}},\ }\href@noop
  {} {\bibfield  {journal} {\bibinfo  {journal} {arXiv:1011.3598}\ } (\bibinfo
  {year} {2010})}\BibitemShut {NoStop}%
\bibitem [{\citenamefont {Kumar}\ and\ \citenamefont
  {Bechhoefer}(2020)}]{avinash}%
  \BibitemOpen
  \bibfield  {author} {\bibinfo {author} {\bibfnamefont {A.}~\bibnamefont
  {Kumar}}\ and\ \bibinfo {author} {\bibfnamefont {J.}~\bibnamefont
  {Bechhoefer}},\ }\bibfield  {title} {\bibinfo {title} {{Exponentially Faster
  Cooling in a Colloidal System}},\ }\href@noop {} {\bibfield  {journal}
  {\bibinfo  {journal} {Nature}\ }\textbf {\bibinfo {volume} {584}},\ \bibinfo
  {pages} {64} (\bibinfo {year} {2020})}\BibitemShut {NoStop}%
\bibitem [{\citenamefont {Ch\'etrite}\ \emph {et~al.}(2021)\citenamefont
  {Ch\'etrite}, \citenamefont {Kumar},\ and\ \citenamefont
  {Bechhoefer}}]{chetrite}%
  \BibitemOpen
  \bibfield  {author} {\bibinfo {author} {\bibfnamefont {R.}~\bibnamefont
  {Ch\'etrite}}, \bibinfo {author} {\bibfnamefont {A.}~\bibnamefont {Kumar}},\
  and\ \bibinfo {author} {\bibfnamefont {J.}~\bibnamefont {Bechhoefer}},\
  }\bibfield  {title} {\bibinfo {title} {{The Metastable {Mpemba} Effect
  Corresponds to a Non-monotonic Temperature Dependence of Extractable Work}},\
  }\href@noop {} {\bibfield  {journal} {\bibinfo  {journal} {Front. Phys.}\
  }\textbf {\bibinfo {volume} {30}},\ \bibinfo {pages} {654271} (\bibinfo
  {year} {2021})}\BibitemShut {NoStop}%
\bibitem [{\citenamefont {Lasanta}\ \emph {et~al.}(2017)\citenamefont
  {Lasanta}, \citenamefont {Vega~Reyes}, \citenamefont {Prados},\ and\
  \citenamefont {Santos}}]{lasanta}%
  \BibitemOpen
  \bibfield  {author} {\bibinfo {author} {\bibfnamefont {A.}~\bibnamefont
  {Lasanta}}, \bibinfo {author} {\bibfnamefont {F.}~\bibnamefont {Vega~Reyes}},
  \bibinfo {author} {\bibfnamefont {A.}~\bibnamefont {Prados}},\ and\ \bibinfo
  {author} {\bibfnamefont {A.}~\bibnamefont {Santos}},\ }\bibfield  {title}
  {\bibinfo {title} {{When the Hotter Cools More Quickly: {M}pemba Effect in
  Granular Fluids}},\ }\href@noop {} {\bibfield  {journal} {\bibinfo  {journal}
  {Phys. Rev. Lett.}\ }\textbf {\bibinfo {volume} {119}},\ \bibinfo {pages}
  {148001} (\bibinfo {year} {2017})}\BibitemShut {NoStop}%
\bibitem [{\citenamefont {Torrente}\ \emph {et~al.}(2019)\citenamefont
  {Torrente}, \citenamefont {L\'opez-Casta\~no}, \citenamefont {Lasanta},
  \citenamefont {Reyes}, \citenamefont {Prados},\ and\ \citenamefont
  {Santos}}]{torrente}%
  \BibitemOpen
  \bibfield  {author} {\bibinfo {author} {\bibfnamefont {A.}~\bibnamefont
  {Torrente}}, \bibinfo {author} {\bibfnamefont {M.~A.}\ \bibnamefont
  {L\'opez-Casta\~no}}, \bibinfo {author} {\bibfnamefont {A.}~\bibnamefont
  {Lasanta}}, \bibinfo {author} {\bibfnamefont {F.~V.}\ \bibnamefont {Reyes}},
  \bibinfo {author} {\bibfnamefont {A.}~\bibnamefont {Prados}},\ and\ \bibinfo
  {author} {\bibfnamefont {A.}~\bibnamefont {Santos}},\ }\bibfield  {title}
  {\bibinfo {title} {{Large {Mpemba}-like effect in a Gas of Inelastic Rough
  Hard Spheres}},\ }\href@noop {} {\bibfield  {journal} {\bibinfo  {journal}
  {Phys. Rev. E}\ }\textbf {\bibinfo {volume} {99}},\ \bibinfo {pages} {060901}
  (\bibinfo {year} {2019})}\BibitemShut {NoStop}%
\bibitem [{\citenamefont {Momp\'o}\ \emph {et~al.}(2021)\citenamefont
  {Momp\'o}, \citenamefont {L{\'o}pez-Casta{\~n}o}, \citenamefont {Torrente},
  \citenamefont {Reyes},\ and\ \citenamefont {Lasanta}}]{mompo}%
  \BibitemOpen
  \bibfield  {author} {\bibinfo {author} {\bibfnamefont {E.}~\bibnamefont
  {Momp\'o}}, \bibinfo {author} {\bibfnamefont {M.}~\bibnamefont
  {L{\'o}pez-Casta{\~n}o}}, \bibinfo {author} {\bibfnamefont {A.}~\bibnamefont
  {Torrente}}, \bibinfo {author} {\bibfnamefont {F.~V.}\ \bibnamefont
  {Reyes}},\ and\ \bibinfo {author} {\bibfnamefont {A.}~\bibnamefont
  {Lasanta}},\ }\bibfield  {title} {\bibinfo {title} {{Memory Effects in a Gas
  of Viscoelastic Particles}},\ }\href@noop {} {\bibfield  {journal} {\bibinfo
  {journal} {Physics of Fluids}\ }\textbf {\bibinfo {volume} {33}},\ \bibinfo
  {pages} {062005} (\bibinfo {year} {2021})}\BibitemShut {NoStop}%
\bibitem [{\citenamefont {Biswas}\ \emph {et~al.}(2020)\citenamefont {Biswas},
  \citenamefont {Prasad}, \citenamefont {Raz},\ and\ \citenamefont
  {Rajesh}}]{rajesh}%
  \BibitemOpen
  \bibfield  {author} {\bibinfo {author} {\bibfnamefont {A.}~\bibnamefont
  {Biswas}}, \bibinfo {author} {\bibfnamefont {V.~V.}\ \bibnamefont {Prasad}},
  \bibinfo {author} {\bibfnamefont {O.}~\bibnamefont {Raz}},\ and\ \bibinfo
  {author} {\bibfnamefont {R.}~\bibnamefont {Rajesh}},\ }\bibfield  {title}
  {\bibinfo {title} {{Mpemba Effect in Driven Granular {Maxwell} Gases}},\
  }\href@noop {} {\bibfield  {journal} {\bibinfo  {journal} {Phys. Rev. E}\
  }\textbf {\bibinfo {volume} {102}},\ \bibinfo {pages} {012906} (\bibinfo
  {year} {2020})}\BibitemShut {NoStop}%
\bibitem [{\citenamefont {Biswas}\ \emph {et~al.}(2022)\citenamefont {Biswas},
  \citenamefont {Prasad},\ and\ \citenamefont {Rajesh}}]{biswas}%
  \BibitemOpen
  \bibfield  {author} {\bibinfo {author} {\bibfnamefont {A.}~\bibnamefont
  {Biswas}}, \bibinfo {author} {\bibfnamefont {V.~V.}\ \bibnamefont {Prasad}},\
  and\ \bibinfo {author} {\bibfnamefont {R.}~\bibnamefont {Rajesh}},\
  }\bibfield  {title} {\bibinfo {title} {{Mpemba Effect in Anisotropically
  Driven Inelastic {Maxwell} Gases}},\ }\href@noop {} {\bibfield  {journal}
  {\bibinfo  {journal} {J. Stat. Phys.}\ }\textbf {\bibinfo {volume} {186}},\
  \bibinfo {pages} {45} (\bibinfo {year} {2022})}\BibitemShut {NoStop}%
\bibitem [{\citenamefont {Jin}\ and\ \citenamefont {Goddard}(2015)}]{jin}%
  \BibitemOpen
  \bibfield  {author} {\bibinfo {author} {\bibfnamefont {J.}~\bibnamefont
  {Jin}}\ and\ \bibinfo {author} {\bibfnamefont {W.~A.~I.}\ \bibnamefont
  {Goddard}},\ }\bibfield  {title} {\bibinfo {title} {{Mechanisms Underlying
  the {Mpemba} Effect in Water from Molecular Dynamics Simulations}},\
  }\href@noop {} {\bibfield  {journal} {\bibinfo  {journal} {J. Phys. Chem. C}\
  }\textbf {\bibinfo {volume} {119}},\ \bibinfo {pages} {2622} (\bibinfo {year}
  {2015})}\BibitemShut {NoStop}%
\bibitem [{\citenamefont {Tao}\ \emph {et~al.}(2017)\citenamefont {Tao},
  \citenamefont {Zou}, \citenamefont {Jia}, \citenamefont {Li},\ and\
  \citenamefont {Cremer}}]{tao}%
  \BibitemOpen
  \bibfield  {author} {\bibinfo {author} {\bibfnamefont {Y.}~\bibnamefont
  {Tao}}, \bibinfo {author} {\bibfnamefont {W.}~\bibnamefont {Zou}}, \bibinfo
  {author} {\bibfnamefont {J.}~\bibnamefont {Jia}}, \bibinfo {author}
  {\bibfnamefont {W.}~\bibnamefont {Li}},\ and\ \bibinfo {author}
  {\bibfnamefont {D.}~\bibnamefont {Cremer}},\ }\bibfield  {title} {\bibinfo
  {title} {{Different Ways of Hydrogen Bonding in Water - Why Does Warm Water
  Freeze Faster than Cold Water?}},\ }\href@noop {} {\bibfield  {journal}
  {\bibinfo  {journal} {J. Chem. Theory Comput}\ }\textbf {\bibinfo {volume}
  {13}},\ \bibinfo {pages} {55} (\bibinfo {year} {2017})}\BibitemShut {NoStop}%
\bibitem [{\citenamefont {Burridge}\ and\ \citenamefont
  {Linden}(2016)}]{burridge}%
  \BibitemOpen
  \bibfield  {author} {\bibinfo {author} {\bibfnamefont {H.~C.}\ \bibnamefont
  {Burridge}}\ and\ \bibinfo {author} {\bibfnamefont {P.~F.}\ \bibnamefont
  {Linden}},\ }\bibfield  {title} {\bibinfo {title} {{Questioning the {Mpemba}
  Effect: Hot Water Does Not Cool More Quickly Than Cold}},\ }\href@noop {}
  {\bibfield  {journal} {\bibinfo  {journal} {Sci. Rep.}\ }\textbf {\bibinfo
  {volume} {6}},\ \bibinfo {pages} {37665} (\bibinfo {year}
  {2016})}\BibitemShut {NoStop}%
\bibitem [{\citenamefont {Lu}\ and\ \citenamefont {Raz}(2017)}]{lu_raz}%
  \BibitemOpen
  \bibfield  {author} {\bibinfo {author} {\bibfnamefont {Z.}~\bibnamefont
  {Lu}}\ and\ \bibinfo {author} {\bibfnamefont {O.}~\bibnamefont {Raz}},\
  }\bibfield  {title} {\bibinfo {title} {{Nonequilibrium Thermodynamics of the
  {Markovian Mpemba} Effect and its Inverse}},\ }\href@noop {} {\bibfield
  {journal} {\bibinfo  {journal} {Proc. Natl. Acad. Sci. U. S. A.}\ }\textbf
  {\bibinfo {volume} {114}},\ \bibinfo {pages} {5083} (\bibinfo {year}
  {2017})}\BibitemShut {NoStop}%
\bibitem [{\citenamefont {Klich}\ \emph {et~al.}(2019)\citenamefont {Klich},
  \citenamefont {Raz}, \citenamefont {Hirschberg},\ and\ \citenamefont
  {Vucelja}}]{klich}%
  \BibitemOpen
  \bibfield  {author} {\bibinfo {author} {\bibfnamefont {I.}~\bibnamefont
  {Klich}}, \bibinfo {author} {\bibfnamefont {O.}~\bibnamefont {Raz}}, \bibinfo
  {author} {\bibfnamefont {O.}~\bibnamefont {Hirschberg}},\ and\ \bibinfo
  {author} {\bibfnamefont {M.}~\bibnamefont {Vucelja}},\ }\bibfield  {title}
  {\bibinfo {title} {{Mpemba Index and Anomalous Relaxation}},\ }\href@noop {}
  {\bibfield  {journal} {\bibinfo  {journal} {Phys. Rev. X}\ }\textbf {\bibinfo
  {volume} {9}},\ \bibinfo {pages} {021060} (\bibinfo {year}
  {2019})}\BibitemShut {NoStop}%
\bibitem [{\citenamefont {Auerbach}(1995)}]{auerbach}%
  \BibitemOpen
  \bibfield  {author} {\bibinfo {author} {\bibfnamefont {D.}~\bibnamefont
  {Auerbach}},\ }\bibfield  {title} {\bibinfo {title} {{Supercooling and the
  {Mpemba} Effect: {When} Hot Water Freezes Quicker Than Cold}},\ }\href@noop
  {} {\bibfield  {journal} {\bibinfo  {journal} {Am. J. Phys.}\ }\textbf
  {\bibinfo {volume} {63}},\ \bibinfo {pages} {882} (\bibinfo {year}
  {1995})}\BibitemShut {NoStop}%
\bibitem [{\citenamefont {Vynnycky}\ and\ \citenamefont
  {Kimura}(2015)}]{vynnycky}%
  \BibitemOpen
  \bibfield  {author} {\bibinfo {author} {\bibfnamefont {M.}~\bibnamefont
  {Vynnycky}}\ and\ \bibinfo {author} {\bibfnamefont {S.}~\bibnamefont
  {Kimura}},\ }\bibfield  {title} {\bibinfo {title} {{Can Natural Convection
  Alone Explain the {Mpemba} Effect?}},\ }\href@noop {} {\bibfield  {journal}
  {\bibinfo  {journal} {Int. J. Heat Mass Transf.}\ }\textbf {\bibinfo {volume}
  {80}},\ \bibinfo {pages} {243} (\bibinfo {year} {2015})}\BibitemShut
  {NoStop}%
\bibitem [{\citenamefont {Zhang}\ \emph {et~al.}(2014)\citenamefont {Zhang},
  \citenamefont {Huang}, \citenamefont {Ma}, \citenamefont {Zhou},
  \citenamefont {Zhou}, \citenamefont {Zheng}, \citenamefont {Jiang},\ and\
  \citenamefont {Sun}}]{xi}%
  \BibitemOpen
  \bibfield  {author} {\bibinfo {author} {\bibfnamefont {X.}~\bibnamefont
  {Zhang}}, \bibinfo {author} {\bibfnamefont {Y.}~\bibnamefont {Huang}},
  \bibinfo {author} {\bibfnamefont {Z.}~\bibnamefont {Ma}}, \bibinfo {author}
  {\bibfnamefont {Y.}~\bibnamefont {Zhou}}, \bibinfo {author} {\bibfnamefont
  {J.}~\bibnamefont {Zhou}}, \bibinfo {author} {\bibfnamefont {W.}~\bibnamefont
  {Zheng}}, \bibinfo {author} {\bibfnamefont {Q.}~\bibnamefont {Jiang}},\ and\
  \bibinfo {author} {\bibfnamefont {C.~Q.}\ \bibnamefont {Sun}},\ }\bibfield
  {title} {\bibinfo {title} {{Hydrogen-bond Memory and Water-skin Supersolidity
  Resolving the {Mpemba} Paradox}},\ }\href@noop {} {\bibfield  {journal}
  {\bibinfo  {journal} {Phys. Chem. Chem. Phys.}\ }\textbf {\bibinfo {volume}
  {16}},\ \bibinfo {pages} {22995} (\bibinfo {year} {2014})}\BibitemShut
  {NoStop}%
\bibitem [{\citenamefont {Tang}\ \emph {et~al.}(2023)\citenamefont {Tang},
  \citenamefont {Huang}, \citenamefont {Zhang}, \citenamefont {Liu},\ and\
  \citenamefont {Zhao}}]{tang}%
  \BibitemOpen
  \bibfield  {author} {\bibinfo {author} {\bibfnamefont {Z.}~\bibnamefont
  {Tang}}, \bibinfo {author} {\bibfnamefont {W.}~\bibnamefont {Huang}},
  \bibinfo {author} {\bibfnamefont {Y.}~\bibnamefont {Zhang}}, \bibinfo
  {author} {\bibfnamefont {Y.}~\bibnamefont {Liu}},\ and\ \bibinfo {author}
  {\bibfnamefont {L.}~\bibnamefont {Zhao}},\ }\bibfield  {title} {\bibinfo
  {title} {{Direct Observation of the {Mpemba} Effect with Water: {Probe} the
  Mysterious Heat Transfer}},\ }\href@noop {} {\bibfield  {journal} {\bibinfo
  {journal} {InfoMat}\ }\textbf {\bibinfo {volume} {5}},\ \bibinfo {pages}
  {e12352} (\bibinfo {year} {2023})}\BibitemShut {NoStop}%
\bibitem [{\citenamefont {Schwarzendahl}\ and\ \citenamefont
  {L\"owen}(2022)}]{lowen22}%
  \BibitemOpen
  \bibfield  {author} {\bibinfo {author} {\bibfnamefont {F.~J.}\ \bibnamefont
  {Schwarzendahl}}\ and\ \bibinfo {author} {\bibfnamefont {H.}~\bibnamefont
  {L\"owen}},\ }\bibfield  {title} {\bibinfo {title} {{Anomalous Cooling and
  Overcooling of Active Colloids}},\ }\href@noop {} {\bibfield  {journal}
  {\bibinfo  {journal} {Phys. Rev. Lett.}\ }\textbf {\bibinfo {volume} {129}},\
  \bibinfo {pages} {138002} (\bibinfo {year} {2022})}\BibitemShut {NoStop}%
\bibitem [{\citenamefont {Ahn}\ \emph {et~al.}(2016)\citenamefont {Ahn},
  \citenamefont {Kang}, \citenamefont {Koh},\ and\ \citenamefont {Lee}}]{ahn}%
  \BibitemOpen
  \bibfield  {author} {\bibinfo {author} {\bibfnamefont {Y.~H.}\ \bibnamefont
  {Ahn}}, \bibinfo {author} {\bibfnamefont {H.}~\bibnamefont {Kang}}, \bibinfo
  {author} {\bibfnamefont {D.~Y.}\ \bibnamefont {Koh}},\ and\ \bibinfo {author}
  {\bibfnamefont {H.}~\bibnamefont {Lee}},\ }\bibfield  {title} {\bibinfo
  {title} {{Experimental Verifications of {Mpemba}-Like Behaviors of Clathrate
  Hydrates}},\ }\href@noop {} {\bibfield  {journal} {\bibinfo  {journal} {Koren
  J. Chem. Eng.}\ }\textbf {\bibinfo {volume} {33}},\ \bibinfo {pages} {1903}
  (\bibinfo {year} {2016})}\BibitemShut {NoStop}%
\bibitem [{\citenamefont {Greaney}\ \emph {et~al.}(2011)\citenamefont
  {Greaney}, \citenamefont {Lani}, \citenamefont {Cicero},\ and\ \citenamefont
  {Grossman}}]{greaney}%
  \BibitemOpen
  \bibfield  {author} {\bibinfo {author} {\bibfnamefont {P.~A.}\ \bibnamefont
  {Greaney}}, \bibinfo {author} {\bibfnamefont {G.}~\bibnamefont {Lani}},
  \bibinfo {author} {\bibfnamefont {G.}~\bibnamefont {Cicero}},\ and\ \bibinfo
  {author} {\bibfnamefont {J.~C.}\ \bibnamefont {Grossman}},\ }\bibfield
  {title} {\bibinfo {title} {{Mpemba-Like Behavior in Carbon Nanotube
  Resonators}},\ }\href@noop {} {\bibfield  {journal} {\bibinfo  {journal}
  {Metall. Mater. Trans. A}\ }\textbf {\bibinfo {volume} {42}},\ \bibinfo
  {pages} {3907} (\bibinfo {year} {2011})}\BibitemShut {NoStop}%
\bibitem [{\citenamefont {Gonz\'alez-Adalid~Pemart\'{\i}n}\ \emph
  {et~al.}(2021)\citenamefont {Gonz\'alez-Adalid~Pemart\'{\i}n}, \citenamefont
  {Momp\'o}, \citenamefont {Lasanta}, \citenamefont {Mart\'{\i}n-Mayor},\ and\
  \citenamefont {Salas}}]{pemartin21}%
  \BibitemOpen
  \bibfield  {author} {\bibinfo {author} {\bibfnamefont {I.}~\bibnamefont
  {Gonz\'alez-Adalid~Pemart\'{\i}n}}, \bibinfo {author} {\bibfnamefont
  {E.}~\bibnamefont {Momp\'o}}, \bibinfo {author} {\bibfnamefont
  {A.}~\bibnamefont {Lasanta}}, \bibinfo {author} {\bibfnamefont
  {V.}~\bibnamefont {Mart\'{\i}n-Mayor}},\ and\ \bibinfo {author}
  {\bibfnamefont {J.}~\bibnamefont {Salas}},\ }\bibfield  {title} {\bibinfo
  {title} {{Slow Growth of Magnetic Domains Helps Fast Evolution Routes for
  Out-of-Equilibrium Dynamics}},\ }\href@noop {} {\bibfield  {journal}
  {\bibinfo  {journal} {Phys. Rev. E}\ }\textbf {\bibinfo {volume} {104}},\
  \bibinfo {pages} {044114} (\bibinfo {year} {2021})}\BibitemShut {NoStop}%
\bibitem [{\citenamefont {Chatterjee}\ \emph {et~al.}(2023)\citenamefont
  {Chatterjee}, \citenamefont {Takada},\ and\ \citenamefont
  {Hayakawa}}]{Hayakawa2023}%
  \BibitemOpen
  \bibfield  {author} {\bibinfo {author} {\bibfnamefont {A.~K.}\ \bibnamefont
  {Chatterjee}}, \bibinfo {author} {\bibfnamefont {S.}~\bibnamefont {Takada}},\
  and\ \bibinfo {author} {\bibfnamefont {H.}~\bibnamefont {Hayakawa}},\
  }\bibfield  {title} {\bibinfo {title} {{Quantum Mpemba Effect in a Quantum
  Dot with Reservoirs}},\ }\href@noop {} {\bibfield  {journal} {\bibinfo
  {journal} {Phys. Rev. Lett.}\ }\textbf {\bibinfo {volume} {131}},\ \bibinfo
  {pages} {080402} (\bibinfo {year} {2023})}\BibitemShut {NoStop}%
\bibitem [{\citenamefont {Biswas}\ \emph {et~al.}(2023)\citenamefont {Biswas},
  \citenamefont {Rajesh},\ and\ \citenamefont {Pal}}]{rajeshmpemba}%
  \BibitemOpen
  \bibfield  {author} {\bibinfo {author} {\bibfnamefont {A.}~\bibnamefont
  {Biswas}}, \bibinfo {author} {\bibfnamefont {R.}~\bibnamefont {Rajesh}},\
  and\ \bibinfo {author} {\bibfnamefont {A.}~\bibnamefont {Pal}},\ }\bibfield
  {title} {\bibinfo {title} {{Mpemba effect in a Langevin system: Population
  Statistics, Metastability, and other Exact Results}},\ }\href@noop {}
  {\bibfield  {journal} {\bibinfo  {journal} {J. Chem. Phys.}\ }\textbf
  {\bibinfo {volume} {159}},\ \bibinfo {pages} {4} (\bibinfo {year}
  {2023})}\BibitemShut {NoStop}%
\bibitem [{\citenamefont {Gr\'egoire}\ and\ \citenamefont
  {Chat\'e}(2004)}]{gregoire04}%
  \BibitemOpen
  \bibfield  {author} {\bibinfo {author} {\bibfnamefont {G.}~\bibnamefont
  {Gr\'egoire}}\ and\ \bibinfo {author} {\bibfnamefont {H.}~\bibnamefont
  {Chat\'e}},\ }\bibfield  {title} {\bibinfo {title} {{Onset of Collective and
  Cohesive Motion}},\ }\href@noop {} {\bibfield  {journal} {\bibinfo  {journal}
  {Phys. Rev. Lett.}\ }\textbf {\bibinfo {volume} {92}},\ \bibinfo {pages}
  {025702} (\bibinfo {year} {2004})}\BibitemShut {NoStop}%
\bibitem [{\citenamefont {Baglietto}\ and\ \citenamefont
  {Albano}(2008)}]{albano08}%
  \BibitemOpen
  \bibfield  {author} {\bibinfo {author} {\bibfnamefont {G.}~\bibnamefont
  {Baglietto}}\ and\ \bibinfo {author} {\bibfnamefont {E.~V.}\ \bibnamefont
  {Albano}},\ }\bibfield  {title} {\bibinfo {title} {{Finite-size Scaling
  Analysis and Dynamic Study of the Critical Behavior of a Model for the
  Collective Displacement of Self-driven Individuals}},\ }\href@noop {}
  {\bibfield  {journal} {\bibinfo  {journal} {Phys. Rev. E}\ }\textbf {\bibinfo
  {volume} {78}},\ \bibinfo {pages} {021125} (\bibinfo {year}
  {2008})}\BibitemShut {NoStop}%
\bibitem [{\citenamefont {Baglietto}\ and\ \citenamefont
  {Albano}(2009)}]{albano2009}%
  \BibitemOpen
  \bibfield  {author} {\bibinfo {author} {\bibfnamefont {G.}~\bibnamefont
  {Baglietto}}\ and\ \bibinfo {author} {\bibfnamefont {E.~V.}\ \bibnamefont
  {Albano}},\ }\bibfield  {title} {\bibinfo {title} {{Nature of the
  Order-Disorder Transition in the {Vicsek} Model for the Collective Motion of
  Self-Propelled Particles}},\ }\href@noop {} {\bibfield  {journal} {\bibinfo
  {journal} {Phys. Rev. E}\ }\textbf {\bibinfo {volume} {80}},\ \bibinfo
  {pages} {050103} (\bibinfo {year} {2009})}\BibitemShut {NoStop}%
\bibitem [{\citenamefont {Landau}\ and\ \citenamefont
  {Binder}(2009)}]{landaubinderbook}%
  \BibitemOpen
  \bibfield  {author} {\bibinfo {author} {\bibfnamefont {D.~P.}\ \bibnamefont
  {Landau}}\ and\ \bibinfo {author} {\bibfnamefont {K.}~\bibnamefont
  {Binder}},\ }\href@noop {} {\emph {\bibinfo {title} {{A Guide to Monte Carlo
  Simulations in Statistical Physics}}}}\ (\bibinfo  {publisher} {Cambridge
  University Press},\ \bibinfo {address} {Cambridge},\ \bibinfo {year}
  {2009})\BibitemShut {NoStop}%
\bibitem [{\citenamefont {Roy}\ and\ \citenamefont {Das}(2013)}]{roydas}%
  \BibitemOpen
  \bibfield  {author} {\bibinfo {author} {\bibfnamefont {S.}~\bibnamefont
  {Roy}}\ and\ \bibinfo {author} {\bibfnamefont {S.~K.}\ \bibnamefont {Das}},\
  }\bibfield  {title} {\bibinfo {title} {{Dynamics and Growth of Droplets Close
  to the Two-Phase Coexistence Curve in Fluids}},\ }\href@noop {} {\bibfield
  {journal} {\bibinfo  {journal} {Soft Matter}\ }\textbf {\bibinfo {volume}
  {9}},\ \bibinfo {pages} {4178} (\bibinfo {year} {2013})}\BibitemShut
  {NoStop}%
\bibitem [{\citenamefont {Majumder}\ and\ \citenamefont
  {Das}(2011)}]{majumder11}%
  \BibitemOpen
  \bibfield  {author} {\bibinfo {author} {\bibfnamefont {S.}~\bibnamefont
  {Majumder}}\ and\ \bibinfo {author} {\bibfnamefont {S.~K.}\ \bibnamefont
  {Das}},\ }\bibfield  {title} {\bibinfo {title} {{Diffusive Domain Coarsening:
  {E}arly Time Dynamics and Finite-size Effects}},\ }\href@noop {} {\bibfield
  {journal} {\bibinfo  {journal} {Phys. Rev. E}\ }\textbf {\bibinfo {volume}
  {84}},\ \bibinfo {pages} {021110} (\bibinfo {year} {2011})}\BibitemShut
  {NoStop}%
\bibitem [{\citenamefont {Pearson}\ and\ \citenamefont
  {Galton}(1895)}]{pearson95}%
  \BibitemOpen
  \bibfield  {author} {\bibinfo {author} {\bibfnamefont {K.}~\bibnamefont
  {Pearson}}\ and\ \bibinfo {author} {\bibfnamefont {F.}~\bibnamefont
  {Galton}},\ }\bibfield  {title} {\bibinfo {title} {{Note on Regression and
  Inheritance in the Case of Two Parents}},\ }\href@noop {} {\bibfield
  {journal} {\bibinfo  {journal} {Proc. R. soc. Lond.}\ }\textbf {\bibinfo
  {volume} {58}},\ \bibinfo {pages} {240} (\bibinfo {year} {1895})}\BibitemShut
  {NoStop}%
\end{thebibliography}%

\end{document}